\newif\iflanl
\openin 1 lanlmac
\ifeof 1 \lanlfalse \else \lanltrue \fi
\closein 1
\iflanl
    \input lanlmac
\else
    \message{[lanlmac not found - use harvmac instead}
    \input harvmac
\fi
\newif\ifhypertex
\ifx\hyperdef\UnDeFiNeD
    \hypertexfalse
    \message{[HYPERTEX MODE OFF}
    \def\hyperref#1#2#3#4{#4}
    \def\hyperdef#1#2#3#4{#4}
    \def\hypernoname{}
    \def\e@tf@ur#1{}
    \def\eprt#1{{\tt #1}}
    \def\CERN{\address{CERN, CH--1211 Geneva 23, Switzerland}}
    \def\wl{W.\ Lerche}
\else
    \hypertextrue
    \message{[HYPERTEX MODE ON}
\def\eprt#1{{\tt #1}}
\def\CERN{\address{

Theory Division, CERN, Geneva, Switzerland}}
\def\wl{

 W.\ Lerche}
\fi
\newif\ifdraft

\noblackbox
\catcode`\@=11
\newif\iffrontpage
\ifx\answ\bigans
\def\titleft{\titla}
\magnification=1200\baselineskip=14pt plus 2pt minus 1pt
%
\advance\hoffset by-0.075truein
\advance\voffset by1.truecm
\hsize=6.15truein\vsize=600.truept\hsbody=\hsize\hstitle=\hsize
\else\let\lr=L
\def\titleft{\titla}
\magnification=1000\baselineskip=14pt plus 2pt minus 1pt
%
\hoffset=-0.75truein\voffset=-.0truein
\vsize=6.5truein
\hstitle=8.truein\hsbody=4.75truein
\fullhsize=10truein\hsize=\hsbody
\fi
\parskip=4pt plus 15pt minus 1pt
%
\newif\iffigureexists
\newif\ifepsfloaded
\def\epsfcheck{
\ifdraft
\input epsf\epsfloadedtrue
\else
  \openin 1 epsf
  \ifeof 1 \epsfloadedfalse \else \epsfloadedtrue \fi
  \closein 1
  \ifepsfloaded
    \input epsf
  \else
\immediate\write20{NO EPSF FILE --- FIGURES WILL BE IGNORED}
  \fi
\fi
\def\epsfcheck{}}
\def\checkex#1{
\ifdraft
\figureexistsfalse\immediate%
\write20{Draftmode: figure #1 not included}
\figureexiststrue
\else\relax
    \ifepsfloaded \openin 1 #1
        \ifeof 1
           \figureexistsfalse
  \immediate\write20{FIGURE FILE #1 NOT FOUND}
        \else \figureexiststrue
        \fi \closein 1
    \else \figureexistsfalse
    \fi
\fi}
\def\missbox#1#2{$\vcenter{\hrule
\hbox{\vrule height#1\kern1.truein
\raise.5truein\hbox{#2} \kern1.truein \vrule} \hrule}$}
\def\lfig#1{
\let\labelflag=#1%
\def\numb@rone{#1}%
\ifx\labelflag\UnDeFiNeD%
{\xdef#1{\the\figno}%
\writedef{#1\leftbracket{\the\figno}}%
\global\advance\figno by1%
}\fi{\hyperref{}{figure}{{\numb@rone}}{Fig.{\numb@rone}}}}
\def\figinsert#1#2#3#4{
\epsfcheck\checkex{#4}%
\def\figsize{#3}%
\let\flag=#1\ifx\flag\UnDeFiNeD
{\xdef#1{\the\figno}%
\writedef{#1\leftbracket{\the\figno}}%
\global\advance\figno by1%
}\fi
\goodbreak\midinsert%
\iffigureexists
\centerline{\epsfysize\figsize\epsfbox{#4}}%
\else%
\vskip.05truein
  \ifepsfloaded
  \ifdraft
  \centerline{\missbox\figsize{Draftmode: #4 not included}}%
  \else
  \centerline{\missbox\figsize{#4 not found}}
  \fi
  \else
  \centerline{\missbox\figsize{epsf.tex not found}}
  \fi
\vskip.05truein
\fi%
{\smallskip%
\leftskip 4pc \rightskip 4pc%
\noindent\ninepoint\sl \baselineskip=11pt%
{\bf{\hyperdef\hypernoname{figure}{{#1}}{Fig.{#1}}}:~}#2%
\smallskip}\bigskip\endinsert%
}

\def\boxit#1{\vbox{\hrule\hbox{\vrule\kern8pt
\vbox{\hbox{\kern8pt}\hbox{\vbox{#1}}\hbox{\kern8pt}}
\kern8pt\vrule}\hrule}}
\def\mathboxit#1{\vbox{\hrule\hbox{\vrule\kern8pt\vbox{\kern8pt
\hbox{$\displaystyle #1$}\kern8pt}\kern8pt\vrule}\hrule}}

%
\font\bigit=cmti10 scaled \magstep1

\font\titla=cmr10 scaled\magstep3
\font\tenmss=cmss10
\font\absmss=cmss10 scaled\magstep1

\newfam\mssfam
\font\footrm=cmr8  \font\footrms=cmr5
\font\footrmss=cmr5   \font\footi=cmmi8
\font\footis=cmmi5   \font\footiss=cmmi5
\font\footsy=cmsy8   \font\footsys=cmsy5
\font\footsyss=cmsy5   \font\footbf=cmbx8
\font\footmss=cmss8
\def\footfont{\def\rm{\fam0\footrm}
\textfont0=\footrm \scriptfont0=\footrms
\scriptscriptfont0=\footrmss
\textfont1=\footi \scriptfont1=\footis
\scriptscriptfont1=\footiss
\textfont2=\footsy \scriptfont2=\footsys
\scriptscriptfont2=\footsyss
\textfont\itfam=\footi \def\it{\fam\itfam\footi}
\textfont\mssfam=\footmss \def\mss{\fam\mssfam\footmss}
\textfont\bffam=\footbf \def\bf{\fam\bffam\footbf} \rm}
\def\tenpoint{\def\rm{\fam0\tenrm}
\textfont0=\tenrm \scriptfont0=\sevenrm
\scriptscriptfont0=\fiverm
\textfont1=\teni  \scriptfont1=\seveni
\scriptscriptfont1=\fivei
\textfont2=\tensy \scriptfont2=\sevensy
\scriptscriptfont2=\fivesy
\textfont\itfam=\tenit \def\it{\fam\itfam\tenit}
\textfont\mssfam=\tenmss \def\mss{\fam\mssfam\tenmss}
\textfont\bffam=\tenbf \def\bf{\fam\bffam\tenbf} \rm}
\ifx\answ\bigans\def\abstractfont{\tenpoint}\else
\def\abstractfont{\def\rm{\fam0\absrm}
\textfont0=\absrm \scriptfont0=\absrms
\scriptscriptfont0=\absrmss
\textfont1=\absi \scriptfont1=\absis
\scriptscriptfont1=\absiss
\textfont2=\abssy \scriptfont2=\abssys
\scriptscriptfont2=\abssyss
\textfont\itfam=\bigit \def\it{\fam\itfam\bigit}
\textfont\mssfam=\absmss \def\mss{\fam\mssfam\absmss}
\textfont\bffam=\absbf \def\bf{\fam\bffam\absbf}\rm}\fi
%
\def\f@@t{\baselineskip10pt\lineskip0pt\lineskiplimit0pt
\bgroup\aftergroup\@foot\let\next}
\setbox\strutbox=\hbox{\vrule height 8.pt depth 3.5pt width\z@}
\def\vfootnote#1{\insert\footins\bgroup
\baselineskip10pt\footfont
\interlinepenalty=\interfootnotelinepenalty
\floatingpenalty=20000
\splittopskip=\ht\strutbox \boxmaxdepth=\dp\strutbox
\leftskip=24pt \rightskip=\z@skip
\parindent=12pt \parfillskip=0pt plus 1fil
\spaceskip=\z@skip \xspaceskip=\z@skip
\Textindent{$#1$}\footstrut\futurelet\next\fo@t}
\def\Textindent#1{\noindent\llap{#1\enspace}\ignorespaces}
\def\foot{\global\advance\ftno by1%
\attach{\hyperref{}{footnote}{\the\ftno}{\footsymbolgen}}%
\vfootnote{\hyperdef\hypernoname{footnote}{\the\ftno}{\footsymbol}}}%
\def\footnote#1{\global\advance\ftno by1%
\attach{\hyperref{}{footnote}{\the\ftno}{#1}}%
\vfootnote{\hyperdef\hypernoname{footnote}{\the\ftno}{#1}}}%
\newcount\lastf@@t           \lastf@@t=-1
\newcount\footsymbolcount    \footsymbolcount=0
\global\newcount\ftno \global\ftno=0
\def\footsymbolgen{\relax\footsym
\global\lastf@@t=\pageno\footsymbol}
\def\footsym{\ifnum\footsymbolcount<0
\global\footsymbolcount=0\fi
{\iffrontpage \else \advance\lastf@@t by 1 \fi
\ifnum\lastf@@t<\pageno \global\footsymbolcount=0
\else \global\advance\footsymbolcount by 1 \fi }
\ifcase\footsymbolcount
\fd@f\dagger\or \fd@f\diamond\or \fd@f\ddagger\or
\fd@f\natural\or \fd@f\ast\or \fd@f\bullet\or
\fd@f\star\or \fd@f\nabla\else \fd@f\dagger
\global\footsymbolcount=0 \fi }
\def\fd@f#1{\xdef\footsymbol{#1}}
\def\space@ver#1{\let\@sf=\empty \ifmmode #1\else \ifhmode
\edef\@sf{\spacefactor=\the\spacefactor}
\unskip${}#1$\relax\fi\fi}
\def\attach#1{\space@ver{\strut^{\mkern 2mu #1}}\@sf}
%
\newif\ifnref
\def\rrr#1#2{\relax\ifnref\nref#1{#2}\else\ref#1{#2}\fi}
\def\ldf#1#2{\begingroup\obeylines
\gdef#1{\rrr{#1}{#2}}\endgroup\unskip}
\def\nrf#1{\nreftrue{#1}\nreffalse}
\def\doubref#1#2{\refs{{#1},{#2}}}
\def\multref#1#2#3{\nrf{#1#2#3}\refs{#1{--}#3}}
\nreffalse

\def\lref{\ldf}

\def\eqn#1{\xdef #1{(\noexpand\hyperref{}%
{equation}{\secsym\the\meqno}%
{\secsym\the\meqno})}\eqno(\hyperdef\hypernoname{equation}%
{\secsym\the\meqno}{\secsym\the\meqno})\eqlabeL#1%
\writedef{#1\leftbracket#1}\global\advance\meqno by1}
\def\eqnalign#1{\xdef #1{\noexpand\hyperref{}{equation}%
{\secsym\the\meqno}{(\secsym\the\meqno)}}%
\writedef{#1\leftbracket#1}%
\hyperdef\hypernoname{equation}%
{\secsym\the\meqno}{\e@tf@ur#1}\eqlabeL{#1}%
\global\advance\meqno by1}
\def\eqnalign#1{\xdef #1{(\secsym\the\meqno)}
\writedef{#1\leftbracket#1}%
\global\advance\meqno by1 #1\eqlabeL{#1}}
%

%
\def\chap#1{\newsec{#1}}
\def\chapter#1{\chap{#1}}
\def\sect#1{\subsec{#1}}
\def\section#1{\sect{#1}}
\def\\{\ifnum\lastpenalty=-10000\relax
\else\hfil\penalty-10000\fi\ignorespaces}
\def\note#1{\leavevmode%
\edef\@@marginsf{\spacefactor=\the\spacefactor\relax}%
\ifdraft\strut\vadjust{%
\hbox to0pt{\hskip\hsize%
\ifx\answ\bigans\hskip.1in\else\hskip .1in\fi%
\vbox to0pt{\vskip-\dp
\strutbox\sevenbf\baselineskip=8pt plus 1pt minus 1pt%
\ifx\answ\bigans\hsize=.7in\else\hsize=.35in\fi%
\tolerance=5000 \hbadness=5000%
\leftskip=0pt \rightskip=0pt \everypar={}%
\raggedright\parskip=0pt \parindent=0pt%
\vskip-\ht\strutbox\noindent\strut#1\par%
\vss}\hss}}\fi\@@marginsf\kern-.01cm}
\def\titlepage{%
\frontpagetrue\nopagenumbers\abstractfont%
\hsize=\hstitle\rightline{\vbox{\baselineskip=10pt%
{\abstractfont\pubnum}}}\pageno=0}
\frontpagefalse
\def\pubnum{}
\def\pdate{\number\month/\number\yearltd}
\def\makefootline{\iffrontpage\vskip .27truein
\line{\the\footline}
\vskip -.1truein\leftline{\vbox{\baselineskip=10pt%
{\abstractfont\pdate}}}
\else\vskip.5cm\line{\hss \tenrm $-$ \folio\ $-$ \hss}\fi}
\def\title#1{\vskip .7truecm\titlestyle{\titleft #1}}
\def\titlestyle#1{\par\begingroup \interlinepenalty=9999
\leftskip=0.02\hsize plus 0.23\hsize minus 0.02\hsize
\rightskip=\leftskip \parfillskip=0pt
\hyphenpenalty=9000 \exhyphenpenalty=9000
\tolerance=9999 \pretolerance=9000
\spaceskip=0.333em \xspaceskip=0.5em
\noindent #1\par\endgroup }
\def\autskip{\ifx\answ\bigans\vskip.5truecm\else\vskip.1cm\fi}
\def\author#1{\vskip .7in \centerline{#1}}

\def\address#1{\ifx\answ\bigans\vskip.2truecm
\else\vskip.1cm\fi{\it \centerline{#1}}}
\def\abstract#1{
\vskip .5in\vfil\centerline
{\bf Abstract}\penalty1000
{{\smallskip\ifx\answ\bigans\leftskip 2pc \rightskip 2pc
\else\leftskip 5pc \rightskip 5pc\fi
\noindent\abstractfont \baselineskip=12pt
{#1} \smallskip}}
\penalty-1000}
\def\endpage{\tenpoint\supereject\global\hsize=\hsbody%
\frontpagefalse\footline={\hss\tenrm\folio\hss}}
\def\ack{\goodbreak\vskip2.cm\centerline{{\bf Acknowledgements}}}
%
%

%
\def\bfone{\relax{\rm 1\kern-.35em 1}}
\def\inbar{\vrule height1.5ex width.4pt depth0pt}
\def\IC{\relax\,\hbox{$\inbar\kern-.3em{\mss C}$}}
\def\ID{\relax{\rm I\kern-.18em D}}
\def\IF{\relax{\rm I\kern-.18em F}}
\def\IH{\relax{\rm I\kern-.18em H}}
\def\II{\relax{\rm I\kern-.17em I}}
\def\IN{\relax{\rm I\kern-.18em N}}
\def\IP{\relax{\rm I\kern-.18em P}}
\def\IQ{\relax\,\hbox{$\inbar\kern-.3em{\rm Q}$}}
\def\IR{\relax{\rm I\kern-.18em R}}
\font\cmss=cmss10 \font\cmsss=cmss10 at 7pt
\def\ZZ{\relax\ifmmode\mathchoice
{\hbox{\cmss Z\kern-.4em Z}}{\hbox{\cmss Z\kern-.4em Z}}
{\lower.9pt\hbox{\cmsss Z\kern-.4em Z}}
{\lower1.2pt\hbox{\cmsss Z\kern-.4em Z}}\else{\cmss Z\kern-.4em
Z}\fi}
\def\a{\alpha}  
 
 \def\l{\lambda}
 
 \def\cB{{\cal B}}
\def\cC{{\cal C}} 

\def\cF{{\cal F}} \def\cG{{\cal G}}
 \def\cI{{\cal I}}
 \def\cK{{\cal K}}
 
\def\cN{{\cal N}}

\def\nup#1({Nucl.\ Phys.\ $\us {B#1}$\ (}
\def\plt#1({Phys.\ Lett.\ $\us  {#1}$\ (}
\def\cmp#1({Comm.\ Math.\ Phys.\ $\us  {#1}$\ (}
\def\prp#1({Phys.\ Rep.\ $\us  {#1}$\ (}
\def\prl#1({Phys.\ Rev.\ Lett.\ $\us  {#1}$\ (}
\def\prv#1({Phys.\ Rev.\ $\us  {#1}$\ (}
\def\mpl#1({Mod.\ Phys.\ Let.\ $\us  {A#1}$\ (}
\def\ijmp#1({Int.\ J.\ Mod.\ Phys.\ $\us{A#1}$\ (}
\def\tit#1|{{\it #1},\ }
%

%

\def\ni{\noindent}
\def\tilde{\widetilde}

\def\us#1{\underline{#1}}

\def\hat{\widehat}

\def\Coeff#1#2{{#1\over #2}}
\def\Coe#1.#2.{{#1\over #2}}
\def\coeff#1#2{\relax{\textstyle {#1 \over #2}}\displaystyle}
\def\coe#1.#2.{\relax{\textstyle {#1 \over #2}}\displaystyle}

\def\to{\rightarrow}
\def\notin{\hbox{{$\in$}\kern-.51em\hbox{/}}}

\def\attac#1{\Bigl\vert
{\phantom{X}\atop{{\rm\scriptstyle #1}}\phantom{X}}}

\def\del{\partial}

\catcode`\@=12

\def\h {{1\over 2}}

\def\ov {\overline}
\def\o {\over}
\def\Li {{\cal L}i}

\def\th {\theta}

\def\Ga {\Gamma}

\def\tr {{\rm Tr}}
\def\det {{\rm det}}

\def\lf {\left}
\def\ri {\right}

\def\re {{\rm Re}}

\def\p {\partial}

\def\lf {\Big}
\def\ri {\Big}

\def\EQN#1#2{$$#2\eqn#1$$}

\def\hm{{z^*}}  
\def\Tau{T}
\def\zet{\xi}
\def\sigm{\varpi}
\def\hg{{\cal C}}
\def\gg{c}
\def\y{{z_S}}
\def\delt{\beta}
\def\appA{A}
\def\appB{B}

\def\elta{{\cal G}}

\def\nihil#1{{\sl #1}}
\def\br{\hfill\break}
\def\np {{ Nucl.\ Phys.} {\bf B}}
\def\pl {{ Phys.\ Lett.} {\bf B}}

\def\prd {{ Phys.\ Rev. D} }


\lref\Fth{
{C.\ Vafa,
 \nihil{Evidence for F theory,}
 Nucl.\ Phys.\ {\bf B469} (1996) 403-418,
 \eprt{hep-th/9602022}; \br}
{D.\  Morrison and C.\ Vafa,
 \nihil{Compactifications of F theory on Calabi-Yau threefolds~I,}
 Nucl.\  Phys.\ {\bf B473} (1996) 74-92,
 \eprt{hep-th/9602114}; \br
 \nihil{Compactifications of F theory on Calabi-Yau threefolds~II,}
 Nucl.\  Phys.\ {\bf B476} (1996) 437-469,
 \eprt{hep-th/9603161}.}
}

\lref\kv{S.\ Kachru and C.\ Vafa, \nihil{Exact results for N=2
 compactifications of heterotic strings,}
 Nucl.\  Phys.\ {\bf B450} (1995) 69-89,
 \eprt{hep-th/9505105}.
}

\lref\KLM{A.\ Klemm, W.\ Lerche and P.\ Mayr,
 \nihil{K3 Fibrations and Heterotic-Type II String Duality,}
 Phys.\  Lett.\ {\bf B357} (1995) 313-322,
 \eprt{hep-th/9506112}.}

\lref\HM{
{J.\ Harvey and G.\ Moore,
 \nihil{Algebras, BPS States, and Strings,}
 Nucl.\  Phys.\ {\bf B463} (1996) 315-368,
 \eprt{hep-th/9510182}.}}

\lref\gm{G.\ Moore, \nihil{String duality,
automorphic forms, and generalized
Kac--Moody algebras}, hep--th/9710198.}

\lref\gins{P.\ Ginsparg,
\nihil{Comment on toroidal compactification of heterotic
superstrings,} \prd {\bf 35} (1987) 648.}

\lref \DKLII{L. Dixon, V. Kaplunovsky and J. Louis,
 \nihil{Moduli dependence of string loop corrections to gauge coupling
constants,} Nucl.\ Phys.\ {\bf B355} (1991) 649-688.}

\lref\msi{P.\ Mayr and S.\ Stieberger,
 \nihil{Threshold corrections to gauge couplings in orbifold
compactifications,}
Nucl.\  Phys.\ {\bf B407} (1993) 725-748,
 \eprt{hep-th/9303017}.}

\lref\fs{K.\ F\"orger and S.\ Stieberger,  {\nihil{String amplitudes
and $N=2$, $d=4$ prepotential in heterotic
$K3\times T^2$ compactifications,}
Nucl.\ Phys.\ {\bf B514} (1998) 135,  \eprt{hep-th/9709004}.}}

\lref\kawai{
{T.\ Kawai,
   \nihil{String duality and modular forms,}
   Phys.\ Lett.\ {\bf B397} (1997) 51-62,
   \eprt{hep-th/9607078};}\br
\nihil{N=2 Heterotic string threshold correction,
K3 surface and generalized Kac--Moody superalgebra},
\pl {\bf 372} (1996) 59, \eprt{hep-th/9512046}.}

\lref\kawainew{T.\ Kawai,
\nihil{String duality and enumeration of curves by Jacobi forms,}
\eprt{hep-th/9804014}.}

\lref\LSW{W.\ Lerche, D.\ Smit and N.\ Warner,
 \nihil{Differential equations for periods and flat coordinates in
two-dimensional topological matter theories,}
 Nucl.\  Phys.\ {\bf B372} (1992) 87-112,
 \eprt{hep-th/9108013}.}

\lref\BK{C.\ Bachas and E.\ Kiritsis,
 \nihil{$F^4$ terms in N=4 string vacua,}
 Nucl.\  Phys.\  Proc.\  Suppl.\ {\bf 55B} (1997) 194,
 \eprt{hep-th/9611205}.}

\lref\BFKOV{
{C.\ Bachas, C.\ Fabre, E.\ Kiritsis, N.\ Obers and P.\ Vanhove,
 \nihil{Heterotic/type I duality and D-brane instantons,}
 Nucl.\  Phys.\ {\bf B509} (1998) 33,
 \eprt{hep-th/9707126}.}
}

\lref\KO{E.\ Kiritsis and N.\ Obers,  \nihil{Heterotic type I duality
in $d <10$-dimensions, threshold corrections and D-instantons,} {\it
JHEP} {\bf 10} (1997) 004, \eprt {hep-th/9709058}.}

\lref\borch{R.E.\ Borcherds, \nihil{Automorphic forms and
$O_{s+2,2}(\IR)$ and infinite products},
{Invent.\ Math.} {\bf 120} (1995) 161; \nihil{Automorphic forms with
singularities on Grassmannians,}
\eprt{alg-geom/9609022}. }

\lref\KDSM{K.\ Dasgupta and S.\ Mukhi,
 \nihil{F theory at constant coupling,}
 Phys.\  Lett.\ {\bf B385} (1996) 125-131,
 \eprt{hep-th/9606044}.}

\lref\LY{
{B.\ Lian and S.\ Yau,
 \nihil{Arithmetic properties of mirror map and quantum coupling,}
 Commun.\  Math.\  Phys.\ {\bf 176} (1996) 163-192,
 \eprt{hep-th/9411234};} \br
{
 \nihil{Mirror maps, modular relations and hypergeometric series 1,}
 \eprt{hep-th/9507151};\nobreak} \br\nobreak
\nobreak{\nobreak
 \nihil{Mirror maps, modular relations and hypergeometric series. 2,}
 \eprt{hep-th/9507153}.}
}

\lref\elias{For a comprehensive review, see: E.\ Kiritsis,
 \nihil{Introduction to nonperturbative string theory,}
 \eprt{hep-th/9708130}.}

\lref\WL{W. Lerche, {
 \nihil{Elliptic index and superstring effective actions,}
 Nucl.\  Phys.\ {\bf B308} (1988) 102.}}

\lref\ellg {A.\ Schellekens and N.\ Warner,
{\nihil{Anomalies, characters and strings,}
 Nucl.\  Phys.\ {\bf B287} (1987) 317;}\br
{E.\ Witten,
 \nihil{Elliptic genera and quantum field theory,}
 Commun.\  Math.\  Phys.\ {\bf 109} (1987) 525;}\br
W. Lerche, B.E.W. Nilsson, A.N. Schellekens and N.P. Warner,
\np {\bf 299} (1988) 91.}

\lref\PADM{
{P.\ Aspinwall and D.\ Morrison,
 \nihil{Point - like instantons on K3 orbifolds,}
 Nucl.\  Phys.\ {\bf B503} (1997) 533,
 \eprt{hep-th/9705104};}\br
{P.\ Aspinwall,
 \nihil{M-theory versus F-theory pictures of the heterotic string,}
 Adv.\  Theor.\  Math.\  Phys.\ {\bf 1} (1998) 127-147,
 \eprt{hep-th/9707014}.}
}

\lref\mirror{
See e.g.,
\nihil{Essays and mirror manifolds}, (S.\ Yau, ed.),
International Press 1992;
\nihil{Mirror symmetry II}, (B.\ Greene et al, eds.),
International Press 1997.
}

\lref\PADMKthree{P.\ Aspinwall and D.\ Morrison,
 \nihil{String theory on K3 surfaces,}
 \eprt{hep-th/9404151}.}

\lref\LS{W.\ Lerche and S.\ Stieberger,
 \nihil{Prepotential, mirror map and F-theory on K3,} 
Adv.\  Theor.\  Math.\  Phys.\ {\bf 2} (1998) 1105-1140; Erratum-ibid.,  
 \eprt{hep-th/9804176}. }

\lref\flohr{M.\ Flohr,
 \nihil{Logarithmic conformal field theory and Seiberg-Witten models,}
 \eprt{hep-th/9808169}.}

\lref\mckay{J.\ Harnad and J.\ McKay,
 \nihil{Modular Solutions to Equations of Generalized Halphen Type,}
 \eprt{solv-int/9804006}.}

\lref\Sen{A.\ Sen,
 \nihil{BPS states on a three brane probe,}
 Phys.\  Rev.\ {\bf D55} (1997) 2501-2503,
 \eprt{hep-th/9608005}.}

\lref\KP{E.\ Kiritsis and B.\ Pioline,
 \nihil{On $R^4$ threshold corrections in IIB string theory
and (p,q) string instantons,}
 Nucl.~ Phys.~{\bf B508} (1997) 509,
 \eprt{hep-th/9707018}.}

\lref\LStW{W.\ Lerche, S.\ Stieberger and N.\ Warner, 
\nihil{Prepotentials from symmetric products}, \eprt{hep-th/9901162}.}

\lref\Fone{M.\ Bershadsky, S.\ Cecotti, H.\ Ooguri, C.\ Vafa,
 \nihil{Holomorphic anomalies in topological field theories,}
 Nucl.~ Phys.~{\bf B405} (1993) 279-304,
 \eprt{hep-th/9302103};\br
\nihil{Kodaira-Spencer theory of gravity and
exact results for quantum string amplitudes,}
 Commun.~ Math.~ Phys.~{\bf 165} (1994) 311-428,
 \eprt{hep-th/9309140}.}

\lref\ZY{S.\ Yau and E.\ Zaslow,
 \nihil{BPS states, string duality, and nodal curves on K3,}
 Nucl.~ Phys.~{\bf B471} (1996) 503-512,
 \eprt{hep-th/9512121}.}

\lref\BL{L. G\"ottsche,
\nihil{A conjectural generating function for numbers of curves on
surfaces,}
\eprt{alg-geom/9711012};\br
L.\ Bryan and N.\ Leung,
 \nihil{The enumerative geometry of K3 surfaces and modular forms,}
 \eprt{alg-geom/9711031}.}

\lref\BVS{M.\ Bershadsky, C.\ Vafa and V.\ Sadov,
 \nihil{D-branes and topological field theories,}
 Nucl.~ Phys.~{\bf B463} (1996) 420-434,
 \eprt{hep-th/9511222}.}

\lref\SW{N.\ Seiberg and E.\ Witten,
 \nihil{Electric-magnetic duality, monopole condensation,
and confinement in $N=2$ supersymmetric Yang-Mills theory,}
 Nucl.\ Phys.\ {\bf B426} (1994) 19-52,
 \eprt{hep-th/9407087}.}

\lref\CCLM{G.\ Cardoso, G.\ Curio, D.\ L\"ust and T.\ Mohaupt,
 \nihil{On the duality between the heterotic string and F-theory in
 eight dimensions,}
 Phys.\  Lett.\ {\bf B389} (1996) 479-484,
 \eprt{hep-th/9609111}.}

\lref\RD{R.\ Dijkgraaf,
 \nihil{Instanton strings and hyperK\"ahler geometry,}
 \eprt{hep-th/9810210}.}

\lref\RDA{R.\ Dijkgraaf,
 \nihil{The Mathematics of five-branes,}
 \eprt{hep-th/9810157}.}

\lref\huy{D.\ Huybrechts,
\nihil{Compact hyperk\"ahler manifolds: basic results},
\eprt{alg-geom/9705025}. }

\lref\FHSV{J.\ Harvey and G.\ Moore,
 \nihil{Exact gravitational threshold correction in the FHSV model,}
 Phys.~ Rev.~{\bf D57} (1998) 2329-2336,
 \eprt{hep-th/9611176}.}

\lref\except{A.N. Redlich, H.J. Schnitzer and K. Tsokos,
\nihil{Bose--Fermi equivalence on the two--dimensional trous for
simply--laced groups}, \np {\bf 289} (1987) 397.}

\lref\ellis{
D.J. Gross, J.A. Harvey, E. Martinec and R. Rohm,
\nihil{Heterotic string theory (2). The interacting heterotic string},
Nucl.\ Phys. {\bf B267} (1986) 75;\br
{W. Lerche, B.E.W Nilsson, A.N. Schellekens and N.P. Warner,
\nihil{Anomaly cancelling terms from the elliptic genus},
Nucl.\ Phys.  {\bf B299} (1988) 91;}\br
{J. Ellis, P. Jetzer and L. Mizrachi, \nihil{One--loop corrections to
the
effective field theory}, Nucl.\ Phys. {\bf B303} (1988) 1.}}

\lref\narain{M.\ Bianchi, E.\ Gava, F.\ Morales and K.\ S.\ Narain,
 \nihil{D--strings in unconventional type I vacuum configurations,}
 \eprt{hep-th/9811013}.
}

\lref\mario{
{E.\ Kiritsis, C.\ Kounnas, P.\ M.\ Petropoulos and J.\ Rizos,
 \nihil{String threshold corrections in models with spontaneously
broken supersymmetry,}
 \eprt{hep-th/9807067};}\br
{A.\ Gregori, C.\ Kounnas and P.\ Petropoulos,
 \nihil{Nonperturbative gravitational corrections in a class of N=2
string duals,}
 \eprt{hep-th/9808024}.}
}

\lref\zagier{M. Eichler and D. Zagier, {\it The theory of Jacobi
forms}, Birkh{\"a}user 1985.}

\lref\stieberg{S.\ Stieberger,
\nihil{(0,2) Heterotic gauge couplings and their $M$--theory origin},
\eprt{hep-th/9807124}, to appear in \np.}

\lref\henning{M.\ Henningson and G.\ Moore,
\nihil{Threshold corrections in $K3\times T^2$ heterotic string
compactifications}, \np {\bf 482} (1996) 187, \eprt{hep-th/9608145}.}

\lref\marino{M.\ Marino and G.\ Moore,
\nihil{Counting higher genus curves in a Calabi--Yau manifold},
\eprt{hep-th/9808131}.}

\lref\ns{H.P.\ Nilles and S.\ Stieberger,
\nihil{String unification, universal one--loop corrections and
strongly coupled heterotic string theory}, \np {\bf 499} (1997) 3,
\eprt{hep-th/9702110}.}

\lref\polch{J. Polchinski,
\nihil{TASI lectures on D-branes},
\eprt{hep-th/9611050}.}

\lref\BS{L.\ Baulieu and S.\ Shatashvili,
 \nihil{Duality from topological symmetry,}
 \eprt{hep-th/9811198}.}

\def\pubnum{
\hbox{CERN-TH/98-378}
\hbox{hep-th/9811228}}
\def\pdate{}
\titlepage
\vskip2.cm
\title
{{\titlefont  Quartic Gauge Couplings from $K3$ Geometry}}
\vskip -.7cm
\autskip
\author{\wl,
S.~Stieberger and N.P.~Warner\footnote{*}{On leave
from Physics Department,
U.S.C., University Park, Los Angeles, CA 90089-0484}}
\vskip0.2truecm
\CERN
\vskip0.2truecm
\vskip-.8truecm

\abstract{
We show how certain $F^4$ couplings in eight dimensions can be computed
using the mirror map and $K3$ data. They perfectly match with the
corresponding heterotic one-loop couplings, and therefore this amounts
to a successful test of the conjectured duality between the heterotic
string on $T^2$ and $F$-theory on $K3$. The underlying quantum geometry
appears to be a 5-fold, consisting of a hyperk\"ahler $4$-fold fibered
over a $\IP^1$ base. The natural candidate for this fiber is the
symmetric product Sym$^2(K3)$. We are lead to this structure by
analyzing the implications of higher powers of $E_2$ in the relevant
Borcherds counting functions, and in particular the appropriate
generalizations of the Picard-Fuchs equations for the $K3$.
}

\vfil
\vskip 1.cm
\ni {CERN-TH/98-378}\hfill\break
\ni November 1998
\endpage
\baselineskip=14pt plus 2pt minus 1pt


\chapter{Introduction}

We consider certain threshold corrections $\Delta(T,U)$ to $F^4$
couplings in eight dimensional string compactifications with $N=1$
supersymmetry. Such theories are obtained from the heterotic string
compactified on $T^2$ (with moduli $T,U$ plus 16 Wilson lines that we
will suppress), or dually, from $F$-theory \Fth\ compactified on
elliptic fibered $K3$'s. Threshold corrections of this kind have been
considered by various authors, either from the heterotic string point
of view or from the dual Type I string perspective
\multref\BFKOV{\KO\LS}\narain. Furthermore, an attempt was made in \LS\
to compute these couplings from $K3$ geometry in $F$-theory; it is the
purpose of the present paper to extend and improve upon this approach.

The motivation for studying this subject is,  of course, not that eight
dimensions would be phenomenologically very important, but rather that
we expect to learn more about how to do exact non-perturbative
computations in $D$-brane physics.\foot{Other interesting aspects of
$D=8$ theories have been recently discussed in \BS.} Experience
suggests that whenever we study BPS-saturated couplings
\multref\WL\HM\BK\ in an effective action, there should be a purely
geometrical method for computing them. Indeed,  we will argue that
there is a beautiful structure behind the 7-brane interactions in eight
dimensions: the relevant quantum geometry appears to be a 5-fold, given
by a fibration of a hyperk\"ahler $4$-fold over a $\IP^1$ base. This
$4$-fold is nothing but the symmetric product Sym$^2(K3)\equiv
{K3\otimes K3\over S_2}$  of the underlying $K3$.

For simplicity, we will focus in this paper only on a certain
class of couplings for one-parameter families of elliptic $K3$'s, and
intend to present a more thorough geometrical treatment in a companion
paper \LStW. We will consider couplings of the form
$$
{\rm Re}\big[\Delta_{G_1G_2}(T)\big]~\Tr\big[F_{G_1}
\wedge F_{G_1}\big]\wedge \Tr\big[F_{G_2}\wedge F_{G_2}\big]\ ,
\eqn\FGGp
$$
where $G_{1,2}$ are non-abelian gauge groups (e.g., $E_8$).
There is no holomorphic prepotential
%
%
underlying this kind of coupling.  Recall that it is only the $U(1)$
couplings of the form ${\rm Re}[{\Delta}_{TTUU}] {F_T}^2{F_U}^2$ etc.\
that possess an underlying holomorphic prepotential, {\it i.e.},
${\Delta}_{TTUU}\sim {\del_T}^2{\del_U}^2{\cal G}(T,U)$ {}~\LS. The
latter class of couplings, and their prepotentials will be discussed in
\LStW.

The situation in eight dimensions is analogous to the more familiar
$N=2$ supersymmetric theories in four dimensions, which are obtained
from the  heterotic string on $K3\times T^2$ and from the type IIA/B
strings on Calabi-Yau 3-folds: there is no holomorphic prepotential,
$\cF$, for couplings of the form ${\rm Re}[\Delta^{{d=4\atop N=2}}
(T,U)]~\Tr[F_{G}\wedge F_{G}]$,\foot{We consider  only the {\it
perturbative} one-loop  piece in four dimensions, and send the  dilaton
to weak coupling, i.e., $e^{-4\pi S}\to 0$. In eight  dimensions the
heterotic one-loop result is supposed to be exact \doubref\BFKOV\KO.}
whereas there is such a prepotential for the couplings of the  $U(1)$
gauge fields $F_{T}$ and $F_{U}$.

More explicitly, the Wilsonian one-loop heterotic string
threshold corrections in four dimensions, after performing the modular
integrations, can be expressed in terms of Borcherds modular products
\nrf{\kawai\henning\ns\gm\fs\marino}\refs{\borch,\HM,\kawai-\marino}
$$
\eqalign{
\Delta^{{d=4\atop N=2}}(T,U)\ &=\ \log[\Psi]\ ,\qquad {\rm where}\cr
\Psi ~&=~ (q_T)^a (q_U)^b ~ \prod_{(k,l)>0} \left(1- {q_T}^k {q_U}^l
\right)^{ \gg(k l)}\ ,}
\eqn\nextwoD
$$
for some $a,b$. Here, $q_T = e^{2 \pi i T}$, $q_U = e^{2 \pi i U}$, the
product runs over $k>0,\ l\in \ZZ\ \ \wedge\ \ k=0, \ l>0$ in the
chamber $T_2\equiv {\rm ImT}>U_2\equiv {\rm ImU}$,  and $\gg(n)$ are
the expansion coefficients of a certain nearly holomorphic and
quasi-modular form,\foot{A modular function is called {\it nearly}
holomorphic if it is meromorphic with poles only at cusps ($\tau=i
\infty$ for $SL(2,\ZZ)$), and we will call such a form {\it
quasi}-modular if it can be written in the form $\hg(\tau) =
P(E_2,E_4,E_6)/\Delta^m$, where $P$ is some (quasi-homogeneous)
polynomial, $\Delta = \eta^{24}(\tau)$, and where $E_n$ are the
familiar Eisenstein functions.} $\hg(\tau)=\sum \gg(n) q^n$. The
precise form of the ``counting function'' $\hg$, depends on the model
and specific gauge group factor that is considered \stieberg.

In spite of the lack of a prepotential, there is a natural geometric
formulation of the  four dimensional couplings $\Delta^{{d=4\atop
N=2}}(T,U)$, and this still involves  the mirror map, and is closely
related to the counting of elliptic curves.\foot{The prepotentials for
the couplings of $F_{T}$ and $F_{U}$ are, of course, related to the
counting of rational curves.} More precisely, the four-dimensional
couplings are sections of a line bundle, which can be trivialized at
large K\"ahler structures using the mirror map $t_k(z_l)$ and the
fundamental period $\varpi_0$. Following an argument given in \FHSV,
$\Psi\,{\rm det}_{kl}({\del t_k\over\del z_l})
{\varpi_0}^{3+h_{1,1}-\chi/12}$ is an invariant ratio of sections,
whose only singularities can be on the discriminant locus of the CY
3-fold. Thus, denoting the components of the discriminant by $D_i$ and
taking the logarithm, we know from this general reasoning that the
couplings can be written in the form:
$$
\Delta^{{d=4\atop N=2}}\ =\
\log\big[\prod_i \Big(D_i(z)^{\a_{i}}\Big)
 {\rm det}_{kl}\lf({\del z_l\over\del t_k}\ri)
{\varpi_0}^{\chi/12-3-h_{1,1}}\big]\ .
\eqn\dfourFone
$$
This has the same form as the topological partition function $\cF_1$
\Fone, which counts elliptic curves in the 3-fold. The couplings
differ from $\cF_1$ in the values of the discriminant exponents $\a_i$,
but we see here the sense in which  the threshold couplings are related
to the counting of elliptic curves. In practice, there   is no easy way
to determine the $\a_i$, other than by matching the asymptotic
behaviour of \dfourFone\ and \nextwoD\ at large K\"ahler structures.

By performing the relevant heterotic one-loop modular integrals
\refs{\BFKOV-\LS}, it turns out that the threshold couplings
$\Delta_{G_1G_2}(T,U)$ in eight dimensions have a product
representation that is completely analogous to the four-dimensional
expression in \nextwoD. One thus may expect that there should be some
way to compute these expressions geometrically, similar in spirit to
\dfourFone. It is the purpose of the present paper to show that this
expectation bears out, by showing that the ${F_{G_1}}^2{F_{G_2}}^2$
threshold corrections can be represented in a way analogous to
\dfourFone\ (where again a few parameters $\a_i$ need to be matched
against the heterotic one-loop result). Our results make major use of,
and indeed generalize the mirror map of the relevant $K3$ surface, and
once again, the threshold corrections are related to counting elliptic
curves in $K3$.

In the next section, we will first analyze the structure of the
relevant Borcherds products that underlie the heterotic one-loop
couplings, for $G_{1,2}=E_8$. The novel feature as compared to the
well-known four-dimensional story is the appearance of ${E_2}^2$ in the
counting functions.  In Section 2.2 we translate this into
properties of the Picard-Fuchs system that the geometrical ($F$-theory)
formulation of the problem must provide. In Section 2.3 we generalize
this to a whole sequence of models with different gauge symmetries,
which have essentially the same structure. In Section 3 we then
interpret the inhomogenous Picard-Fuchs equations of Section 2.2 in
terms of geometry, and are thereby naturally lead to symmetric products
of $K3$ and their fibrations.

Finally, in appendix \appA\ we discuss some properties of quasi-modular
Borcherds products, while in appendix B we present a streamlined
technique for the computation of the heterotic one-loop couplings. Here
we also show that these couplings can be obtained concisely in terms of
a generating function that has an intriguing interpretation in terms of
$D$-strings. 

\chapter{Borcherds products and mirror map}

\section{Building blocks}

To simplify the discussion, we focus here on the model with $E_8\times
E_8$ non-abelian gauge symmetry and fixed modulus $U=\rho\equiv e^{2\pi
i/3}$; we will later show how our arguments can easily be
generalized to a whole series of one-parameter models.

An algebraic representation of the relevant singular $K3$
with two $E_8$ singularities is given by
$$
W(x,y,\zet)\ =\
y^2 + x^3 + \zet^5(\zet-1)(\zet-{\hm}(\tau)) \ =\ 0\ .
\eqn\kthree
$$
The mirror map, namely the map to the flat coordinate $T$, is
\doubref\CCLM\LS
$$
\hm(T)\ =\ \big(\sqrt{-j(T)/1728}+\sqrt{1-j(T)/1728}\big)^2\ ,
\eqn\mirmap
$$
which is nothing but the hauptmodul for a
certain $\ZZ_2$ extension of the modular group, $SL(2,\ZZ)$. It
is more convenient for our work to use $SL(2,\ZZ)$ modular forms,
and so we introduce
$$
z(T)\ \equiv\ -4 {\hm(T)\over (1-\hm(T))^2} \ =\ {1728\over j(T)}\ .
\eqn\ztdef
$$

Our task is to represent the $F^4$ heterotic threshold corrections, as
computed \KO\ and rederived in appendix B.2, in terms of the mirror map
pertaining to the $K3$ surface \kthree. The product form of these
looks exactly like \nextwoD, but with $U=\rho$:
$$
\eqalign{
\Delta_{E_8E_8'}(T) = - 48\, \log[\Psi]\attac{U\equiv\rho}
 \ , &\quad {\rm with}~
a=-2\ , b= 0 \ , {\rm and\ counting\ function}\cr & \hg~=~{1\over
12}~{1\over
\eta^{24}}
\Big[ E_2 E_4 - E_6 \Big]^2
\cr
\Delta_{E_8E_8}(T) = - 24\, \log[\Psi]\attac{U\equiv\rho}
 \ , &\quad {\rm with}~
a= 8\ , b= 12 \ , {\rm and\ counting\ function}\cr &  \hg~=~{1\over
12}~
{E_4\over \eta^{24}}~\Big[ E_2^2 E_4 - 2 E_2  E_6 + E_4^2 \Big] \ .
\cr
}\eqn\gEE
$$
We see (due to the finite number of quasi-modular forms with a given
degree) that all couplings are composed out of a finite number of
building blocks. Most importantly, note that there are two kinds of
ingredients: \item{$(i)$} those terms that are fully modular and are
polynomials in the Eisenstein series $E_4(\tau)$ and $E_6(\tau)$
\item{$(ii)$} those terms that are quasi-modular, because they
involve powers of $E_2(\tau)$.

The theorems of Borcherds \borch\ state that the product $\Psi(T,U)$
in \nextwoD\  has good modular properties essentially if $\hg(\tau)$ is
a modular function\foot{If the counting function, $\hg$, has weight
zero then the constant term of its $q$-expansion is required to be
divisible by $24$.}.  However these theorems do not apply if $\hg$
contains $E_2$.  This means that while the pieces of \gEE\ that do not
contain $E_2$ map into the ring of modular functions generated by
$z(T)$, $z(T)-1$ and $z(T)'$, the $E_2$-parts cannot map into
this ring.  However, as we will see in the next section and in appendix
\appA, we can make good use of the fact that the $E_2$ pieces arise
from taking derivatives of true modular forms (and that the $E_2$
pieces can be removed from the counting function by further judicious
differentiation). In this spirit, we parametrize the non-modular pieces
in the following way:
$$
\eqalign{
 &\hg_1~\equiv~{1 \over 2 \pi i} {d \over d \tau} {E_4  E_6
\over \eta^{24}} ~=~ - {1 \over \eta^{24} }\left( {1 \over 2} E_4^3
+ {1 \over 3} E_6^2 + {1 \over 6} E_2 E_4 E_6 \right)\ , \cr
&\hg_2~\equiv~ \left({1 \over 2 \pi i} {d \over d \tau}\right)^2
{E_4^2  \over \eta^{24}} ~=~ - {1 \over \eta^{24} }\left( {13 \over 36}
E_4^3 + {2 \over 9} E_6^2 + {1 \over 3} E_2 E_4 E_6 +
{1 \over 12} E_2^2 E_4^2 \right)\ , }
$$
and define
$$
\mu_i(T)\ =\ a_i~\log\Big[q_T^{-1} ~
\prod_{(k,l)>0} \left(1- {q_T}^k {q_U}^l
\right)^{\gg_i(k l)}\attac{U\equiv\rho}\!\!\!\Big]\ ,
\qquad i=1,2.\eqn\mumodprod
$$
where $a_i$ are some normalization constants that will be fixed later
($a_1=-3,\,a_2=-9/2$). Combining this with the modularly well-behaved
pieces, we can now rewrite the threshold couplings in terms of these
building blocks in the following way:
$$
\Delta(T)\ =\ -48\big(\log\big[ z(T)^{\a_1} (z'(T))^{\a_2}
(z(T)-1)^{\a_3}\big]
 + \delt_1 \mu_1(T)  + \delt_2 \mu_2(T)\big)\ .
\eqn\building
$$
Explicitly, comparing with \gEE, we find that\foot
{Note that $\Delta_{E_8E_8'}-2 \Delta_{E_8E_8}=288 \log[\eta(T)^{24}]$,
which represents the eight dimensional analog of the well-known
result \DKLII\ about differences of four dimensional
threshold couplings.}
$$
\eqalign{
\Delta_{E_8E_8'}: &\qquad  {\a_1}=-2,\, {\a_2}=0,\,
{\a_3}=0,\,\delt_1=-1,\,\delt_2=2/9, \cr \Delta_{E_8E_8}: & \qquad
{\a_1}=-16,\, {\a_2}=18,\, {\a_3}=-9,\,\delt_1=-1/2,\,\delt_2=1/9.
\cr}\eqn\gEE
$$

Equation \building\ is the analogue, and in fact the generalization of
the threshold formula \dfourFone\ in four dimensions. Indeed the
corresponding four-dimensional expression can be written exactly in
this form, but with $\delt_2=0$. Note that the four-dimensional
expression \dfourFone\  is modular as a function of {\it all} the
Calabi-Yau moduli, including the dilaton modulus $\y\sim e^{-4\pi S}$.
The lack of modularity (due to the $\mu_1$) comes from identifying the
perturbative coupling, $S$, and extracting the weak coupling limit. The
non-modular function $\mu_1-\log(z) $ then turns up as the finite
residue in $\lim_{S \to \infty} (\log(\y) - S)$.  (The $\log(z)$ term
subtracts the singularity at $T=U=\rho$.)

The vanishing of $\delt_2$ is a reflection of the fact that for the
four-dimensional gauge couplings, $E_2$ appears only linearly in the
counting function, and not quadratically. The new feature in eight
dimensions is thus the presence of the function $\mu_2$, whose
Borcherds formula has a counting function containing  $E_2^2$. This
raises the question as to how such functions would naturally appear
from the intrinsic geometry of $K3$.  In fact, counting functions of
curves of algebraic genus $g$ with $n$-nodes, passing through $g$
points on $K3$, have been found in \BL:
$$
\hg_g\ \equiv\ \sum_{n=0}^\infty \gg_g(n)q^n\ =\
\Big({\del\over \del q}E_2(q)\Big)^g {q\over \eta(q)^{24}}
\ .
$$
These involve arbitrary high powers of $E_2$, and in particular
one has:
$
\hg_1={1 \over 12} {E_2^2-E_4\over\eta^{24}}
$.
This means that the threshold corrections in eight dimensions can
formally be
related to the counting of nodal elliptic curves in $K3$.

\section{Picard-Fuchs equations with sources}

We now wish to relate the functions $\mu_i$ to the geometry of the dual
$F$-theory: that is, to the geometry of the relevant elliptically
fibered $K3$ \kthree. In practice this means that we want to obtain a
generalization of the usual Picard-Fuchs operator. At $U=\rho$, this PF
operator is of second order, and after transforming to the variable
$z(T)$ in \ztdef, it becomes:
$$
{\cal L}^{(2)}  ~\equiv~ {1 \over z}~
\Big[ \theta_z^2 ~-~ z~(\theta_z + {5 \over 12})(\theta_z +
{1 \over 12}) \Big]\ ,
\eqn\kthreePF
$$
where $\theta_z \equiv z {d \over d z}$.  The fundamental solutions to
${\cal L}^{(2)} \sigm_i(z) =0$ are given by the periods
$$
\sigm_0 (z) ~=~ {}_2F_1({1 \over 12},{5 \over 12};1,z)~=~ (E_4)^{1/4}
\ , \quad \sigm_1 (z) ~=~ T~\sigm_0 ~=~ T~(E_4)^{1/4}\ .
\eqn\kthreeperiods
$$
As was noted in \LY, there is a canonical association
of \kthreePF\ to the following third-order operator:
$$
{\cal L}^{(3)}  ~\equiv~ {1 \over z}~
\Big[ \theta_z^3 ~-~ z~(\theta_z + {5 \over 6})(\theta_z +
{1 \over 2})(\theta_z +  {1 \over 6}) \Big] \ ,
\eqn\thirdorder
$$
The two operators ${\cal L}^{(2)}$ and ${\cal L}^{(3)}$ are naturally
related with one another for a number of reasons.  First, their
fundamental solutions are quadratically related:
$$
\omega_j (z) ~=~ \sigm_{j-i}~\sigm_{i} ~=~ T^j~(E_4)^{1/2}\ ,
\qquad j=0,1,2,
\eqn\quadrel
$$
where ${\cal L}^{(3)} \omega_i(z) =0$.
This fact will be important later when we discuss the
interpretation of the underlying geometry.

More generally, these two operators satisfy some
interesting identities when filtered through the mirror
map: for any function $f(z)$ one has
\EQN\intidents{\eqalign{z~{\cal L}^{(2)}~(f(z)\sigm_0(z)) ~=~
&{1 \over E_4(q_T)}~(\theta_{q_T}^2~f(z(q_T))) ~\sigm_0\ , \cr
z~{\cal L}^{(3)}~(f(z)\omega_0(z)) ~=~
&{1 \over E_6(q_T)}~(\theta_{q_T}^3~f(z(q_T))) ~\omega_0\ .}}
{}From this and (A.6) it follows that the functions
we seek, $\mu_j(z)$, satisfy the following inhomogenous, or
``source'' PF equations:
\EQN\inhomeqns{
\eqalign{{\cal L}^{(2)}~(\mu_1~\sigm_0(z)) ~&=~
 \sigm_0\  \cr  {\cal L}^{(3)}~(\mu_2~
\omega_0(z)) ~&=~   \omega_0\ ,}}
where we have fixed the normalization constants,
$a_1=-3,\,a_2=-9/2$, in \mumodprod\ by requiring ``unit sources''
on the right-hand sides of these equations.
The solutions of these equations are ambiguous up to additions
of the homogeneous solutions, which amount to
irrelevant addition of terms linear in $T$ to $\mu_1$
and up to quadratic terms in $T$ to $\mu_2$.

Amongst other things, these equations mean that at $U=\rho$ the
Borcherds products $\mu_i$ become solutions to relatively simple linear
systems of equations. In particular, note that ${\cal L}^{(2)}({\cal
L}^{(2)} (\mu_1~\sigm_0(z))) = 0 $ and ${\cal L}^{(3)}({\cal L}^{(3)}
(\mu_2~\omega_0(z))) = 0 $. In other words, we find that the
ingredients $\mu_i$ in the threshold corrections \building\ satisfy
generalized hypergeometric equations of fourth and sixth order,
respectively.

 The question arises as to the physical and geometrical interpretation
of the inhomogenous Picard-Fuchs equations \inhomeqns. We derived them
by working backwards, {\it i.e.}, by investigating  how to reproduce
the threshold corrections originally obtained from the heterotic
string.  However, before we discuss the physical and geometric
interpretation, we first wish to generalize our ideas to a larger class
of models.

\section{Generalization to certain one-parameter families of $K3$'s.}

Consider the sequence of models that have been introduced in ref.\ \LS.
They represent certain one-parameter\foot{The $T^2$ modulus $U$, as
well as the Wilson lines, are frozen to particular finite values~\LS.}
families of singular $K3$ surfaces, with the special property that the
modulus $\tau_s$ of the elliptic fiber (the type IIB string coupling)
remains constant over the base $\IP^1$. These families can be
represented by the following polynomial equations $W(x,y,\zet)=0$:
$$
\eqalign{
({E_8}^2{H_0}^2)&:\ \ \ y^2 + x^3 + \zet^5(\zet-1)(\zet-{\hm}(\Tau)) \
\ \ \,=\ 0 \cr ({E_7}^2{H_1}^2)&:\ \ \ y^2 + x^3 + x
\zet^3(\zet-1)(\zet-{\hm}(\Tau)) \ \ =\ 0 \cr ({E_6}^2{H_2}^2)&:\ \ \
y^2 + x^3 + \zet^4(\zet-1)^2(\zet-{\hm}(\Tau))^2 \ =\ 0 \cr
({D_4}^4)&:\ \ \ y^2 + x^3 + \zet^3(\zet-1)^3(\zet-{\hm}(\Tau))^3 \,=\
0.\cr
}\eqn\Kthrees
$$
The first model is exactly the model with $E_8\times E_8$ gauge
symmetry that we discussed above. Each of these models has four
singularities in the $z$--plane of the indicated types, leading to
corresponding gauge symmetries in $D=8$ (the Kodaira singularities of
type $H_n$ lead to gauge groups $A_n$). There exist actually further
models of the same kind, which we will not discuss in great detail in
the following (but which could be treated in a similar way). That is,
the list of one-parameter families with constant coupling and four
singularities in the $z$-plane includes also the models
$(E_8H_0{D_4}^2)$, $(E_7H_1{D_4}^2)$, $(E_6H_2{D_4}^2)$,
$(E_8H_0E_6H_2)$, $({E_6}^2D_4H_0)$ and $({H_2}^2D_4E_8)$.

One feature these models have in common is that their mirror maps are
uniformly given by certain Thompson series; this is much in line of
the findings of ref.~\LY. The abovementioned models indeed match very
well with the list of replicable arithmetic triangle functions
discussed in \mckay. More specifically, explicit computations show that
the mirror maps are determined by the Schwarzian equation
$$
{{{\hm}}'''\over {{\hm}}'}-{3\over2}({{{\hm}}''\over
{{\hm}}'})^2\ =\ -2Q({{\hm}})\,{{{\hm}}'}^2\ ,
\eqn\swar
$$
where
$$
Q({\hm})\ =\ {1\over4}\left\{{{1-\l^2\over {{\hm}}^2} + {1-\mu^2\over
({\hm}-1)^2}
+{\l^2+\mu^2-\nu^2-1\over {{\hm}}({\hm}-1)}}\right\}\ .
\eqn\Qdef
$$
The solution of \swar\ is given by the Schwarzian triangle function
$$
\Tau({\hm})\ =\ s(\l,\mu,\nu; {\hm})\ ,
\eqn\triangle
$$
where $(\pi\l,\pi\mu,\pi\nu)$ are the angles of the relevant
fundamental domain (which depends on the specific model). We list these
and other data, partly taken from \mckay, in Table~1.

\def\th{\theta}
{\vbox{\ninepoint{ $$ \vbox{\offinterlineskip\tabskip=0pt
\halign{\strut\vrule# &~$#$~\hfil &~$#$~\hfil &~$#$~\hfil &~$#$~\hfil
&~$#$\hfil &\vrule# \cr \noalign{\hrule} & {\rm Elliptic} & {\rm
constant\ IIB} & {\rm Angles} & {\rm Hypergeometric} & {\rm Inverse\
mirror\ map\ =} &\cr & {\rm Singularities} & {\rm coupling}\ \tau_s &
(\l,\mu,\nu) & {\rm indices\ }(a,b;c) & {\rm Hauptmodul\ {\hm}(\Tau)}
&\cr \noalign{\hrule} & {E_8}^2{H_0}^2 & \rho & (0,\coeff23,0) &
(\coeff16,\coeff16,1) & (\sqrt{-J(\Tau)}+\sqrt{1-J(\Tau)})^2 &\cr &
{E_7}^2{H_1}^2 & i & (0,\coeff12,0) & (\coeff14,\coeff14,1) &
-\coeff1{64}(\coeff{\eta(\Tau)}{\eta(2\Tau)})^{24} &\cr &
{E_6}^2{H_2}^2 & \rho & (0,\coeff13,0) & (\coeff13,\coeff13,1) &
-\coeff1{27}(\coeff{\eta(\Tau)}{\eta(3\Tau)})^{12} &\cr & {D_4}^4 &
{\rm any} & (0,0,0) & (\coeff12,\coeff12,1) &
-\coeff1{16}(\coeff{\eta(\Tau)}{\eta(4\Tau)})^8 &\cr & E_8H_0{D_4}^2 &
\rho & (0,\coeff13,\coeff13) & (\coeff16,\coeff12,1) & {\sqrt{3}i\left(
 \th_2^4(2\Tau)-e^{\pi i/3}\th_3^4(2\Tau)\right)^3\over
9\th_2^4(2\Tau)\th_3^4(2\Tau)\th_4^4(2\Tau)}\!-\!1 &\cr & E_7H_1{D_4}^2
& i & (0,\coeff14,\coeff14) & (\coeff14,\coeff12,1) &
{i\left(\th_3^2(2\Tau) + i \th_4^2(2\Tau)\right)^4\over
8\th_2^4(2\Tau)\th_3^2(2\Tau)\th_4^2(2\Tau)} \!-\!1 &\cr &
E_6H_2{D_4}^2 & \rho & (0,\coeff16,\coeff16) & (\coeff13,\coeff12,1) &
{\sqrt{3}i \left(\eta^6(2\Tau) + 3\sqrt{3}i\eta^6(6\Tau)\right)^2\over
36\eta^6(2\Tau)\eta^6(6\Tau)}\!-\!1\!\! &\cr & E_8H_0E_6H_2 & \rho &
(0,\coeff12,\coeff16) & (\coeff16,\coeff13,1) & &\cr & {\displaystyle{
{E_6}^2D_4H_0 \atop {H_2}^2D_4E_8 }}\Big\} & \rho &
(\coeff16,\coeff13,\coeff16) & (\coeff16,\coeff13,\coeff56) & &\cr
\noalign{\hrule}} \hrule}$$ \vskip-10pt \noindent{\bf Table 1:} {\sl
Complete list of one-parameter families of $K3$ surfaces with four
elliptic singularities and constant coupling. The triple $(\l,\mu,
\nu)$ describes the angles of the fundamental region of the relevant
triangle group,  and $(a,b;c)$ the  indices of the corresponding
hypergeometric equation. Every vanishing angle corresponds to a cusp
and thus to a decompactification limit Im$\Tau\to\infty$; the last two
models obviously do not have such a limit ($J\equiv j/1728$). }
\vskip10pt}}

Note that for these models all monodromies (induced by encircling the
four singularities in the $z$-plane) are of finite order. As was
discussed in \LS, this means that the geometry of the singular $K3$'s
can be described by a finite covering of the $z$-plane and thus
effectively reduces to the one of Riemann surfaces; the four $7$-planes
then correspond to the branch points of these curves. More
specifically, for the four models in \Kthrees\ one finds the following
$\ZZ_N$-symmetric curves
$$
\Sigma_N\,:\qquad x^N \ =\ \zet^{-1}(\zet-1)(\zet-\hm)
\eqn\zncurves
$$
of genus $g=N-1$, where $N=6,4,3,2$, respectively.  Indeed, the
relevant period integrals $\varpi_i=\int_{\gamma_i}\!\!dx\, d
\xi/(\del_y W(x,y,\xi))$ of the $K3$ surfaces \Kthrees\ can be directly
obtained from the curves \zncurves. This can be seen by changing
variables in the integral by setting $x = v ~\xi^{2 (1 -
1/N)}(\xi-1)^{2/ N} (\xi - z^*)^{2 /N}$, upon which the integral then
factorizes into:
$
\int  {d v \over \sqrt{v^3 + 1}}\int
{d\zet\over \zet^{1-1/N}(\zet-1)^{1/N}(\zet-{\hm}(\Tau))^{1/N}}.
$
The integral over $v$ is simply a constant normalization,
and we thus reduce the relevant $K3$ periods
to the periods of the $\ZZ_N$ curves:
$$
\sigm_i\ =\ \int
{d\zet\over \zet^{1-1/N}(\zet-1)^{1/N}(\zet-{\hm}(\Tau))^{1/N}}\ .
$$
This can also be interpreted \LS\ as integrals over open string
metrics \Sen, $d\zet \prod_{i=1}^{24}(\zet-\zet_i)^{-1/12}$.
The periods may be written as hypergeometric functions
$$
\eqalign{
\sigm_0\ &=\  (-1)^{-2/N} \pi\, {\rm csc}(\pi/N)\,
       {}_2F_1\big(1/N,1/N,1; {\hm} \big) \cr
\sigm_1\ &= {{\hm}}^{-1/N} (-1)^{-2/N} \pi\, {\rm csc}(\pi/N)\,
{}_2F_1\big(1/N,1/N,1; 1/{\hm} \big)\ .
}\eqn\curveper
$$
of the corresponding $(a,b;c)$ type, as indicated in Table~1.
The flat coordinate is then alternatively given by
$\Tau=\sigm_1/\sigm_0$.

The issue is to compute couplings of the form
$\Delta_{G_1G_2}(\Tau){F_{G_1}}^2{F_{G_2}}^2$ \FGGp, where $G_{1,2}$
are the non-abelian gauge groups of any two given 7-planes, out of the
total of four. As discussed in \LS, the primary, and potentially
singular contribution to this coupling comes from integrating out the
exchange of the $RR$ four-form tensor field $C^{(4)}$ between the two
given 7-planes, simply because each of the planes carries a
world-volume coupling of the form $C^{(4)}\wedge F_{G_i}\wedge
F_{G_i}$.

 It was proposed in \LS\ that the coupling should be given by a
logarithmic correlation function between the two relevant branch points
(7-planes) of $\Sigma_N$.\foot{We suspect that this can be naturally
expressed in terms of a ``logarithmic'' conformal field theory, along
the lines of ref.~\flohr.} This correlator is supposedly nothing but
the Green's function $\elta^{\Sigma_N}$ between appropriate
$1/N$-period points of a scalar field on $\Sigma_N$, ie.,
$\Delta_{G_1G_2}\sim \elta^{\Sigma_N}(\zet_1,\zet_2)$.

The problem is that a Green's function is not uniquely defined since
there is the freedom of adding a non-singular piece to it,
$\elta^{\Sigma_N}(\zet_1,\zet_2,\Tau) \to
\elta^{\Sigma_N}(\zet_1,\zet_2,\Tau)+\delt_i\mu_i(\Tau)$. The canonical
choice for it, given by the prime form, turns out not to give the
complete result in general. More precisely, somewhat tedious explicit
computations show that the canonical Green's function between any two
relevant branch points $z_i$ is composed out of the Hauptmodul $\hm(T)$
and has the general form
$$
\elta^{\Sigma_N}_{\rm{prime\atop form}}(\zet_1,\zet_2,\Tau)\ =\
\log\big[\hm^{\a_1}(1-\hm)^{\a_2}(\hm')^{\a_3}\big]
\eqn\hauptgreen
$$
for an appropriate choice of ${\a_i}$ (this is essentially a
combination of the generalized Halphen functions discussed in
ref.~\mckay.) We find that this Green's function yields the correct
result for the couplings \FGGp\ only for the model with ${D_4}^4$ gauge
symmetry, as was shown in ref.~\LS.\foot{Note that the heterotic loop
computation in \LS\ missed a term, which slighly modifies the result
given in \LS; however, the correlators can be still represented in the
form \hauptgreen\ with the choice:
$(\alpha_1,\alpha_2,\alpha_3)=(1,-1,0),\ (-1,-1,0),\ (-1,1,0)$  
referring to $\Delta_{12}, \Delta_{13}, \Delta_{14}(T)$, respectively.
The correct computation can be found in a seperate erratum.}

The point is that the prime form \hauptgreen\ describes only the
``modular'' part of the threshold correction, but misses the functions
$\mu_i$  in \building.   Physically, \hauptgreen\ describes only the
tree-level exchange of $C$ fields, but misses certain instanton
contributions. Namely, loops of $(p,q)$ strings in the $\zet$-plane
will be closed in general only on the covering surface $\Sigma_N$, so
that such strings effectively wrap the Riemann surfaces. Wrapping
entire world-sheets of such strings will thus in general generate extra
instanton-like contributions. In the ${D_4}^4$ model considered in \LS\
 there are no such instanton corrections ($\delt_i=0$) because
$\Sigma_2$ has genus $g=1$, so that from the point of view of the
$(p,q)$ instantons the situation is like a type IIB compactification on
$T^2$ with maximal supersymmetry: it is known \KP\ that for this
compactification there are no $(p,q)$ instanton corrections to
parity-odd couplings.

The functions $\mu_i$ to be added to the canonical Greens functions
\hauptgreen\ can be obtained in exactly the same way as we did before.
We first perform a quadratic change of variables,\foot {For
$E_{8-k}H_k{D_4}^2$ ($k=0,1,2$), the transformation is $z(T)= -
{\hm(T)^2\over 4 (1-\hm(T))}$, which maps to the equations (2.22) and 
(2.24). For $E_8H_0E_6H_2$, we have simply
$z(T)=\hm(T)$ which maps to these equations for $N=3$. For the last two
entries in Table 1, the transformation (2.21) maps to hypergeometric
systems of types ${}_2F_1(1/12,1/4;5/6,z)$ and
${}_3F_2(1/6,1/2,1/3;2/3,5/6,z)$, respectively.}
$$
z(T)\ \equiv\ -4 {\hm(T)\over (1-\hm(T))^2} \ ,
\eqn\sameztdef
$$
in terms of which the Picard-Fuchs operators are:
$$
{\cal L}^{(2)}_N  ~\equiv~ {1 \over z}~
\Big[ {\theta_z}^2 ~-~ z~(\theta_z + {1 \over 2N})(\theta_z +
{1 \over 2}- {1 \over 2N}) \Big]
\eqn\kthreePFN
$$
where $N=2,3,4,6$, respectively. The fundamental solutions to
${\cal L}^{(2)}_N \sigm_i(z) =0$ are
$$
\eqalign{
&\sigm_0 (z) ~=~ {}_2F_1({1 \over 2N},{1 \over 2}- {1 \over 2N};1,z)\
=\
\sqrt {z' z^{-1}(1-z)^{-1}}\cr
&\sigm_1 (z) ~=~ T~\sigm_0 \ .
}\eqn\kthreeperiodsN
$$
The third-order operators that are associated with \kthreePFN\ are
simply \LY
$$
{\cal L}^{(3)}_N  ~\equiv~ {1 \over z}~
\Big[ {\theta_z}^3 ~-~ z~(\theta_z + 1-{1 \over N})(\theta_z +
{1 \over 2})(\theta_z +  {1 \over N}) \Big] \ ,
\eqn\thirdorderN
$$
whose solutions are again quadratic in terms of $\sigm_i$:
$\omega_j (z) = \sigm_{j-i}~\sigm_{i}$.
We can then analogously write down the source equations:
$$
\eqalign{{\cal L}^{(2)}_N~(\mu_1~\sigm_0(z)) ~&=~
 \sigm_0\cr  {\cal L}^{(3)}_N~(\mu_2~
\omega_0(z)) ~&=~   \omega_0\ ,}
\eqn\sourceN
$$
which finally determine the extra contributions, $\mu_i(z(T))$. Once
again, for simplicity we have chosen to normalize the $\mu_i$ to
satisfy these equations with ``unit source''.

In order to test our ideas explicitly, we now consider the remaining
models in the list \Kthrees, i.e., the ones with $(E_6\times A_2)^2$
and $(E_7\times A_1)^2$ gauge symmetry, and compare the geometric data
with the heterotic one-loop couplings (these one-loop couplings are
are computed in appendix \appB). Since these models have a greater
variety of non-abelian group factors than the $E_8\times E_8$ and
${D_4}^4$ models, there are more couplings to test.

The upshot is that we indeed find that the generic expression
\building\ reproduces the heterotic one-loop results, provided that we
choose the coefficients ${\a_i},\delt_i$ appropriately (where, of
course, $\hm(T)=-\coeff1{27}(\eta(\Tau)/\eta(3\Tau))^{12}$ or
$\hm(T)=-\coeff1{64}(\eta(\Tau)/\eta(2\Tau))^{24}$, respectively, and
where $\mu_{1,2}$ are the solutions of \sourceN\ with $N=3,4$).
Explicitly, by matching the asymptotic $q$-expansions of these building
blocks with the heterotic couplings (B.26) at $U=\rho-1$ \LS,
we have for the $E_6$ model:
$$
\eqalign{
\Delta_{E_6{E_6}'}(T)\ &=\ \log\big[\hm^{-1/3}(\hm-1)^{2/3}\big]
-\coeff1{12} \mu_1 + \coeff1{108} \mu_2 \cr
\Delta_{E_6{A_2}}(T)\ &=\ \log\big[\hm^{-1/6}(\hm-1)^{-1/3}\big]
 +\coeff1{108}\mu_2 \cr
\Delta_{E_6{A_2}'}(T)\ &=\ \log\big[\hm^{-1/3}(\hm-1)^{-1/3}\big]
+\coeff1{108}\mu_2 \cr
\Delta_{A_2{A_2}'}(T)\ &=\ \log\big[\hm^{-1/3}(\hm-1)^{2/3}\big]
+\coeff1{12} \mu_1 + \coeff1{108}\mu_2 \ .\cr
}\eqn\esix
$$
Quite similarly, for the $E_7$ model we find that at $U=1+i$:
$$
\eqalign{
\Delta_{E_7{E_7}'}(T)\ &=\ \log\big[\hm^{-1/12}(\hm-1)^{1/6}\big]
-\coeff1{32} \mu_1 + \coeff1{192} \mu_2 \cr
\Delta_{E_7{A_1}}(T)\ &=\ \log\big[\hm^{-1/24}(\hm-1)^{-1/12}\big]
 +\coeff1{192}\mu_2 \cr
\Delta_{E_7{A_1}'}(T)\ &=\ \log\big[\hm^{-1/12}(\hm-1)^{-1/12}\big]
+\coeff1{192}\mu_2 \cr
\Delta_{A_1{A_1}'}(T)\ &=\ \log\big[\hm^{-1/12}(\hm-1)^{1/6}\big]
+\coeff1{32} \mu_1 + \coeff1{192}\mu_2 \ .\cr
}\eqn\eseven
$$
Thus, including the results of \LS\ and of Section 2.1, we have
verified  that for all $K3$ surfaces in \Kthrees\ we can match the
geometric data to the corresponding heterotic one-loop results. This
represents, we believe, the most complete quantitative test of the
heterotic $F$-theory duality to date.

\chapter{Interpretation and Discussion}

We have demonstrated that the inhomogenous Picard-Fuchs equations
\sourceN\  carry the relevant information about the $F^4$ couplings
\FGGp. We now give two interpretations of these equations.

The first is to note that the structure of the inhomogenous
Picard-Fuchs equations is highly reminiscent of the equations of
Seiberg and Witten \refs{\SW}. Indeed, the geometry of the specific
families \Kthrees\  of singular elliptic $K3$'s effectively reduces to
the one of $SU(N)$ SW curves.  More generally, remember that the
periods, $a$ and $a_D$, of the Seiberg-Witten differential satisfy a
first order system of differential equations:
\EQN\SWeqns{{\del \over \del \hm}~a_D ~=~ \sigm_1 \ , \qquad
{\del \over \del \hm}~a  ~=~ \sigm_0 \ ,}
where the functions $\sigm_i$ are the standard
periods \curveper\ of the $\ZZ_N$ curves \zncurves.

For our quartic gauge couplings in eight dimensions
it is not first order, but second order
operators ${\cal L}^{(2)}_N$ whose application yields the standard
periods of the curves. This means that $\mu_1$ may be seen as a
period of another meromorphic differential on these Riemann surfaces.
Similarly, since the differential operators ${\cal L}^{(3)}_N$ are the
PF operators associated \LY\ with the $K3$ manifolds $X_{6}(1,1,1,3)$,
$X_{4}(1,1,1,1)$, $X_{2,3}(1,1,1,1,1)$ and $X_{2,2,2}(1,1,1,1,1,1)$,
respectively, this suggests that one could associate $\mu_2$ with the
periods of certain meromorphic differentials on these $K3$ surfaces.

A second, and more directly useful interpretation can be given for the
second order equation in \sourceN\ for $\mu_1$, and this will then help
us to get a better understanding of the third-order equation.

As mentioned earlier, the function $\mu_1$ naturally appears also in
the four dimensional, $N=2$ supersymmetric theories arising from 3-fold
compactifications of type II strings. This function is essentially the
difference of $\log(\y) - S$ in the large base space limit of the
relevant Calabi-Yau $3$-fold (in which the non-perturbative
contributions to the threshold corrections drop out).  The relevant
3-folds are known to be $K3$ fibrations \KLM\ over a $\IP^1$ base, and
this implies that the Picard-Fuchs operators of these Calabi-Yau
manifolds must involve, in some way, the differential operators ${\cal
L}^{(2)}_N$ in \kthreePFN.

More precisely, the ``fibered'' PF operators are obtained, to leading
order in $\y\sim e^{-4\pi S}$, by the replacement  $\theta_z^2 \to
\theta_z( \theta_z -2 \theta_\y)$ in the first term of ${\cal
L}^{(2)}_N$.  If one now recalls that $S\sigm_0 \sim (\log(\y)  + \mu_1
- \log(z)) \sigm_0$ is a period of the Calabi-Yau
manifold and if one keeps all the finite terms in the Calabi-Yau
Picard-Fuchs system  in the limit as $S \to \infty$, one finds that
$\theta_\y (\log(\y) \sigm_0)$ contributes a finite term that may be
written as a ${\cal L}^{(2)} ((\mu_1 - \log(z))\sigm_0) = 2
\theta_z \sigm_0$. This equation then trivially reduces to  \sourceN.

In other words, the source term  of the inhomogenous second order
equation \sourceN\ is nothing but a remnant of the heterotic dilaton in
the large base space, or weak coupling limit.

This suggests a natural interpretation of the third order equation
\sourceN,  which appears only for the eight dimensional, but not for
the four dimensional couplings.  A crucial insight can be gained by
paying attention to the structure of the solutions of the {\it
homogenous} equation,  ${\cal L}^{(3)}_N \omega_i(z) =0$: the three
solutions are nothing but quadratic products of the ordinary $K3$
periods. We believe that these periods are to be interpreted as those
of the symmetric product, Sym$^2(K3)$, of the underlying $K3$.

The appearance of Sym$^2(K3)$ is indeed quite natural in the context of
$D$-brane  physics. That is, the contribution to the couplings \FGGp\
we consider comes from pairs of 7-branes, and a system of two branes
(or points on $K3$)  is thought to be described by a non-linear
sigma-model whose target space is Sym$^2(K3)$ \BVS. Since this is a
hyperk\"ahler manifold,\foot {For a review and references, see \huy\
(and also \doubref\RD\RDA).}  and a sigma-model on such a space has
$N=(4,4)$ supersymmetry, the quantum cohomology is trivial and this is
exactly what is reflected by the product structure of the periods.

More generally, any hyperk\"ahler manifold has a holomorphic
$(2,0)$-form and a holomorphic $(4,0)$-form (which may be thought of as
the square of the $(2,0)$-form). It is the variation of the Hodge
structure of the holomorphic $(2,0)$-form and $(4,0)$-form that seems
to underly our two functions $\mu_1$ and $\mu_2$. More precisely, what
we should have is a fibration of these forms, which --in the large base
limit-- manifests itself in the source terms of the inhomogenous
equation \sourceN. The $\IP^1$ fibration yields in total a 5-fold, and
indeed it was suggested in \LS\ a $5$-fold should underlie the $F^4$
couplings in eight dimensions.

We have made extensive, and thus far unsuccessful, attempts  to obtain
algebraic (hyperk\"ahler-fibered) 5-folds, whose Picard-Fuchs systems
would reduce to the source equations presented in this paper. However,
it is notoriously difficult to find algebraic descriptions of
hyperk\"ahler manifolds \huy, and so our lack of success may merely be
reflection of this fact.

The question whether the threshold corrections described in this paper
can indeed be realized in terms of a fibration of Sym$^2(K3)$ or not,
has potentially important physical significance. Remember that what we
just have been arguing is that the heterotic one-loop couplings are
given by the large base space limit of this fibration, just as for the
well-known couplings in four dimensions. However, in four dimensions
this is not the full story, in that the expansion away from the large
base space limit gives the dilaton dependent, non-perturbative
corrections to the one-loop couplings.

One may thus be tempted to ask for an interpretation of the higher
orders of expansion in the base-space parameter, $\y$, of the 5-fold.
It has been suggested \doubref\BFKOV\KO, however, that the heterotic
one-loop corrections to $F^4$ are exact in eight dimensions and that
there are no further non-perturbative corrections. If this were true,
then the source equations discussed in this paper would indeed capture
the complete story.   However, being related to a singular geometrical
limit, this seems a little unnatural;  perhaps there is, in fact, a
physically meaningful extra dependence on a geometrical modulus which
perturbs away from the singular limit. In fact, it is known that
Sym$^2(K3)$ has an extra modulus that controls the blow up of its
$\ZZ_2$ singularity,\foot{It deforms Sym$^2(K3)$ to a smooth Hilbert
scheme \RD.}
$$
{\rm dim}H^{1,1}\big({\rm Sym}^2(K3)\big)\ =\ {\rm dim}H^{1,1}(K3)+1
\ =\ 21\ ,
\eqn\extraH
$$
and it is a non-trivial fact \RDA\ that this modulus behaves
exactly like a string coupling constant. We hope to give a more
detailed presentation of these matters elsewhere.

\ack

We would like to thank
L.\ Alvarez-Gaum\'e,
P.\ Candelas,
R.\ Donagi,
S.\ Ferrara,
B.\ Pioline,
and especially P.\ Mayr for valuable discussions. Furthermore, we like
to thank J.\ McKay for drawing attention to ref.\ \mckay.

\appendix{\appA}{Quasi-modular Borcherds products}

It was shown by Borcherds that if the
``counting function'' $\hg(\tau)\equiv \sum \gg(n)q^n$ is
a true modular form of weight $-s/2$, then there
is a canonical choice of the exponents $a,b$ in
\EQN\borcherd{\Psi ~=~ (q_T)^a (q_U)^b ~ \prod_{(k,l)>0}
\left(1- {q_T}^k {q_U}^l \right)^{ \gg(k l)} \ ,}
such that $\Psi$ is a meromorphic modular form of
$(T,U)$-weights $(\gg(0)/2,\gg(0)/2)$. Moreover, the
zeroes and poles of $\Psi$ are given
precisely by the vanishing of the various factors
in the product. Perhaps the most familiar example of these
Borcherds formulae is:   $\hg(\tau) = E_4^3/\Delta - 744$
then for $T_2 > U_2$ one has $a=-1, b=0$, and
\EQN\borchjfn{\Psi_0 ~=~  j(T) - j(U)\ .}

We want  to find some form of generalized Borcherds
formulae for simplifying modular products involving
$E_2$.  Counting functions involving $E_2$ can be obtained
by differentiating modular polylogarithms.  That is,
consider
\EQN\polylog{\chi(T,U) ~=~ \left({1 \over 2 \pi i}
\right)^{2n+1}~ \sum_{(k,l)>0}~\gg(k l)~\Li_{2n+1}\big[{q_T}^k
{q_U}^l \big]\ ,}
where the polylogarithm is defined by $(a \geq 1)$:
\EQN\LiDefn{\Li_a(z)=\sum_{p>0} {z^p \over p^a}\ , \qquad
{\rm with}\ \Big(z{\del\over\del z}\Big)^{a-1} \Li_a(z)=
-log( 1-z) \ ,}
and, as usual, the sum in \polylog\ runs over the positive
roots $k>0,\ l\in \ZZ\ \ \wedge\ \ k=0,\ l>0$.
It then follows that if one defines $\Psi$ by taking
$\log(\Psi)= (- {1 \over 4 \pi^2 }\del_U \del_T)^n \chi $,
then $ \Psi $ has counting
function $({1 \over 2 \pi i} \del_\tau)^n \hg(\tau)$.

The issue is that the obvious modular quantity \polylog\ has
polylogarithmic singularities, while the natural meromorphic object,
$\Psi$, is not modular.  However one can find a meromorphic, modular
object by further differentiating $\log(\Psi)$. It is elementary to
show that if $F^{(-2 m)}(\tau)$ is a modular form of weight $- 2m$,
then $G^{(2m+2)} = {d^{2m+1} \over  d \tau^{2m+1}} F^{(-2 m)}$ is a
modular form of weight $2m+2$.  That is, $G^{(2m+2)}$ contains {\it no}
$E_2$'s.  Moreover ${d^{m} \over d \tau^{m}} F^{(-2 m)}$ is an
quasi-modular function that contains a factor of $E_2^m F^{(-2 m)}$.
Thus not only is it most natural to think of any $E_2$ in $\hg(\tau)$
as coming from derivatives of other modular forms, but one can render
such functions modular once again by taking a suitable number of
derivatives. For example, define:
\EQN\FGprods{\eqalign{\Psi_i &~=~ q_T^{-1} ~
\prod_{(k,l)>0} \left(1- {q_T}^k {q_U}^l
\right)^{\gg_i(k l)}\ ,\ i=1,2 \cr
 &\hg_1~\equiv~{1 \over 2 \pi i} {d \over d \tau} {E_4  E_6
\over \eta(\tau)^{24}} ~=~
- {1 \over \eta(\tau)^{24} }\left( {1 \over 2} E_4^3
+ {1 \over 3} E_6^2 + {1 \over 6} E_2 E_4 E_6 \right)\ , \cr
& \hg_2~\equiv~ \left({1 \over 2 \pi i} {d \over d \tau}\right)^2
\!\!{E_4^2  \over \eta(\tau)^{24}} ~=~
 {-1 \over \eta(\tau)^{24} }\left( {13 \over 36}
E_4^3 + {2 \over 9} E_6^2 + {1 \over 3} E_2 E_4 E_6 +
{1 \over 12} E_2^2 E_4^2 \right). }}
One can easily check that $({1 \over 2 \pi i}\del_\tau)^2
\hg_1 = (984 - j(\tau))E_4$ and $({1 \over 2 \pi i}\del_\tau)^3
\hg_2 = (240 - j(\tau))E_6$, which
are modular forms of weight $4$ and $6$ respectively.

Now consider  $\chi_1 = \del_U^2 \del_T^2 log(\Psi_1)$ and $\chi_2 =
\del_U^3 \del_T^3log( \Psi_2)$.  These may be viewed as modular
``polylogarithms'' of the form \polylog\ with $n=-2$ and $n=-3$.  The
functions $\Li_a$ for $a \le 0$ are rational, and indeed the
corresponding modular ``polylogarithms'' are the positive weight
automorphic forms generated via the Hecke transformations, and are thus
nearly holomorphic modular forms \borch. The weights of these modular
``polylogarithms'' is the same as the weight of the counting function,
and so the functions $\chi_1$ and $\chi_2$ have $(T,U)$ weight $(4,4)$
and $(6,6)$ respectively. One can use this, and the manifest zeroes and
poles of $\Psi_i$ to uniquely identify the $\chi_i$.  We will not do
this here, but instead focus on the special point $U=\rho\equiv e^{2\pi
i/3}$.

Since we are taking $U =\rho$, we will only be interested in the
modular and holomorphic properties as a function of $T$.  We therefore
consider the $\mu_j =  a_j \log(\Psi_j)|_{U= \rho}$ with the constants
$a_j$ as in \mumodprod, and define $\Phi_j= ({1 \over 2 \pi i}
\del_T)^{j+1}~ \mu_j$, $j=1,2$. The function $\Phi_j$ is thus a modular
form  of weight $2(j+1)$. {}From the product formula \FGprods, and the
fact that $\hg_i \sim -{1 \over q} + const+ \dots$, one sees that the
the functions $\log(\Psi_i)$ are {\it only} singular at $T=U=\rho$, and
moreover, at this point $\Phi_1$ and $\Phi_2$ have double and triple
poles respectively.  One can also easily see that $\Phi_1$ and $\Phi_2$
both vanish at $T = i \infty$.  This determines  the functions $\Phi_i$
up to overall normalizations, and the latter can be fixed by using the
fact that $E_4(\tau)/E_6(\tau) \sim -{2 \pi i \over 3}(\tau - \rho)$ as
$\tau \to \rho$ and using \FGprods\ to obtain the coefficient of the
pole in $\Phi_j$.  One needs to be a little careful in that the product
in \FGprods\ has a simple zero at $T=U$, but in the limit $U \to \rho$
this becomes a triple zero because $U=\rho$ is a $\ZZ_3$-orbifold point
of the fundamental domain.  One finds:
\EQN\Phieqns{\Phi_1 ~=~ 1728~{E_4(q_T) \over j(q_T)} \ , \qquad
\Phi_2 ~=~ 1728 ~{E_6 (q_T) \over j(q_T)} \ . }
%

\appendix{\appB}{Elliptic genera and heterotic $F^4$--corrections}

In this appendix we compute the one-loop threshold corrections in the
heterotic string picture. They are needed in Section 2 for the
comparison with the geometric $K3$ data. In subsection B.1 we will
first write a compact generating functional, from which these couplings
can be obtained by differentiation and which has an interesting
$D$-string interpretation. In B.2 we consider the model with ${E_8}^2$
gauge symmetry, and in B.3 we extend this to the remaining models with
$[E_7\times SU(2)]^2$ and $[E_6\times SU(3)]^2$ gauge symmetry.
Finally, in B.4 we collect some data on Jacobi forms.

\section{1/2 BPS--saturated $F^n$--amplitudes}

We will present here a formal expression for heterotic one--loop
corrections to  $\tr F^n, (\tr F^{n/2})^2, \ldots$\  (in general
$n$--derivative) gauge couplings ($n$=even), where the  gauge fields
originate from $E_8 \times E_8'$ and where the trace is taken in the
adjoint representation. Furthermore, we restrict to $T^2\times X$
heterotic string compactifications and 1/2-BPS saturated amplitudes.
The latter restriction guarantees that the whole left--moving fermionic
part of the partition function (supplemented with $2n$ fermionic
zero modes) cancels against the left-moving
bosonic oscillator contribution. This leads to a world--sheet torus
integral whose integrand is essentially the product of the torus
partition function $Z_{2,2}(T,U)$ and the holomorphic genus
$\Phi_{-n}(q,y)$.  More precisely, we have
$$
\Delta_{(\tr F_{E_8}^{n/2})^2}={1\o (2\pi i)^n}{\p^n \o \p \ov  z^n}
\lf.\int {d^2\tau \o \tau_2}\ [Z_{2,2}(q,\ov q)
\ \tilde{\Phi}_{-n}(\ov q,\ov y)-c_{(n/2)}(0)]\ri|_{\ov z=0}
\ ,\eqn\todo
$$
where $\tilde{\Phi}_{-n}(q,y) =e^{m\pi {z^2\o \tau_2}}\Phi_{-n}(q,y)$
with $y=e^{2\pi i z}$ and $q=e^{2\pi i \tau}$. 
As usual, the non-harmonic pieces are needed
for modular invariance and come from the coincidence of external gauge
legs.   The parameter, $y$, represents one of the skew eigenvalues of
the background gauge field, $F$.   Here we simplify our calculations
(without loss of generality) by restricting attention to a single
such parameter. The constant $c_{(n/2)}(0)$ in \todo\ is defined to be
$E_2^{n/2}(q)\Phi_{-n}(q,1)|_{{\rm coeff}(q^0)}$ and is needed
to keep the integral IR--finite.  The $\Phi_{-n}(q,y)$ are Jacobi
functions with weight $-n$ and index $m=4$, and we define their
expansion coefficients $c(k,b)$ by:
$$
\Phi_{-n}(q,y) =\sum_{k\geq 0}\sum_{b^2 \leq 4mk} c(k,b) y^b q^k\ .
\eqn\exp
$$

In contrast to (B.28), the function $\tilde\Phi_{-n}(q,y)$
has a well-behaved transformation behaviour:
$$
\tilde\Phi_{-n}\lf({a\tau +b \o c\tau+d},{z\o c\tau+d}\ri)=
(c\tau+d)^{-n}\tilde\Phi_{-n}(\tau,z)\ .
\eqn\enjo
$$
It is this property that allows to use in \todo\  the orbit
decompositon method of \DKLII, and after some work to eventually
arrive at (for the chamber $T_2>U_2$ and regularization
$\epsilon\rightarrow \infty$)
$$
\kern-1.5em\eqalign{\Delta_{(\tr F_{E_8}^{n/2})^2}
\!\!\!\!\!\!\!\!\!\!\!\!\!\!\!\!\!\!
&\qquad\ \ \ =
{1\o (2\pi i)^n}{\p^n \o \p z^n}\ \times\cr
&\ \ \lf\{\Big[\!\sum_b\!\sum_{(k,l)>0}
\sum_{p>0}{2\o\sqrt{p^2-{mz^2\o T_2U_2}}}\,
e^{-2\pi(kT_2+lU_2)\sqrt{p^2-{mz^2\o T_2U_2}}}e^{2\pi ip(kT_1+lU_1)}\
c(kl,b)y^b+hc.\Big]\ri.\cr
&\lf.\lf.+\sum_b\lf[{U_2\o \pi}\sum_{j>0}{2\o j^2-mz^2{U_2\o T_2}}+
\sum_{j>0}\lf({2\o\sqrt{j^2-{mz^2\o T_2U_2}}}-{2\o\sqrt{j^2-{mz^2\o
T_2U_2}+
{\epsilon\o \pi T_2U_2}}}\ri)\ri]c(0,b)y^b\ri\}\cr
&+\lf.{\pi\o 3}T_2\sum_{s=0}^{n/2}{1\o s+1}E_2^{s+1}
F_s\ri|_{{\rm coeff}(q^0)}-
c_{(n/2)}(0)[\ln\epsilon+\gamma_E+1+\ln({2\o 3\sqrt 3})]
\,\ri|_{z=0}\ ,\cr}
\eqn\RESULT
$$
with ${3\o 2^{n/2}}\lf.(y {\p \o \p y})^n\tilde
\Phi(q,y)\ri|_{z=0}=:\sum\limits_{s=0}^{n/2} \hat{E_2}^sF_s$. Note that
the last four terms give simply polynomials in $T_2$ and $U_2$.

The formula \RESULT\ can be easily generalized to combinations
$\tr F_1^{n/2}{\tr F_2}^{n/2}$ of different gauge groups, by including
further Wilson lines $z_i\equiv \tr F_i$ and differentiating with
respect to them ($z^2\rightarrow \sum_i z_i^2,\ y^b\rightarrow \prod_i
y_i^{b_i}$).

The complicated formula \RESULT\ has an intriguing physical
interpretation in term of the dual Type I string picture of the
heterotic string, by recognizing the exponentiated square root as a
Born-Infeld action (this generalizes the observations of \BFKOV).
Specifically, in eight dimensions where $n=4$, \RESULT\ can be
rewritten in terms of the Born--Infeld action of a $D$--string, which
reads \polch:
$$
S_{BI}[\cG,\cB,\cF,\cC_2]=
\int d^2\sigma e^{-\phi}\sqrt{\det(\cG+\cB+\cF)}-i\int \cC_{2}\ ,
\eqn\BI
$$
where $\cF=\pmatrix{0&f\cr -f&0\cr}$ is the open string world--volume
$U(1)$ gauge background field. Moreover, in \BI\ we also have the
induced moduli fields $\cG_{\alpha\beta}= G_{ij}\p_\alpha X^i \p_\beta
X^j$,  $\cB_{\alpha\beta}=B_{ij}\p_\alpha X^i \p_\beta X^j$ 
(in what follows $\cB=0$) and the $RR$ $2$--form $\cC_2$ on the world 
volume.  The sum
$k>0,\ l\in \ZZ$ in \RESULT\ over the heterotic winding states thus can
be seen as the D--instanton sum, so that
$$
\Delta_{(\tr F^2)^2}=\lf.
{\p^4 \o \p f^4}\sum
{1\o \sqrt{\det(\cG+\cF)}}
\ e^{-S_{BI}[\cG,\cF,\cC_2]}\
\Phi_{-4}({\cal U},\sqrt{\det\cF})\ri|_{\cF=0}\ ,
\eqn\bi
$$
with the D--brane complex structure
${\cal U}= {j+pU_1\o k}+{p\o k}U_2\sqrt{{\det(\cG+\cF)\o \det\cG}}$,
gauge field $e^{-\phi}f=iz k\sqrt{m{T_2\o U_2}}$, 
$e^{-\phi}\sqrt{\det \cG}=kpT_2$ and $\cC_2=kpT_1$.
On the other hand, the part of \RESULT\ that does not involve
winding states $(k=0)$ gives the perturbative contributions
in Type I language \BFKOV.

We now apply the generating function in \RESULT\ to the three physical
models that we discuss in the present paper.

\section{Gauge group $E_8 \times E_8$ }

Literally taken, the expression for $\Delta_{\tr F_{E_8}^n}$
in \todo\ directly applies to heterotic compactifications on: $(i)$\
$K3\times T^2$ (for $n=2$), or $(ii)$\ $T^2$ (for $n=4$). Indeed, using
$$
\eqalign{
\lf.  (y{\p \o \p y})^2J_{E_8}(q,y)\ri|_{z=0}&=
\lf. 4(y{\p \o \p y})^2E_{4,1}(q,y)\ri|_{z=0}={2\o 3} (E_2E_4-E_6),\cr
\lf.  (y{\p \o \p y})^4J_{E_8}(q,y)\ri|_{z=0}&={4\o 3}
(E_2^2E_4-2E_2E_6+E_4^2)\ .\cr}
\eqn\dif
$$
we can immediately rederive from \RESULT\ the results of \HM\ and
\KO:\br

$(i)$\ $F^2$ in $d=4$, with $\Phi_{-2}(q,y)=
{E_6J_{E_8}(q,y)\o\eta^{24}}$:
$$
\eqalign{\Delta_{\tr F_{E_8}^2}&=
4\re\lf\{\!\!\sum_{(k,l)>0}c_{(1)}(kl)\Li_1(x)-{3\o \pi
T_2U_2}c(kl)\lf[(kT_2+lU_2)
\Li_2(x)+{1\o 2\pi}\Li_3(x)\ri]\ri\}\cr
&-c_{(1)}(0)\ln({\cal K}T_2U_2)-{\pi c(0)\o 15}{U_2^2\o
T_2}-{3c(0)\zeta(3)\o
\pi^2T_2U_2}+{\pi\o 3}c_{(1)}(0)U_2+288\pi T_2\ ,\cr}
\eqn\Four
$$
with ${3\o 2^{s/2}}(y{\p\o \p y})^s\Phi_{-n}(q,y)\lf.\ri|_{z=0}=\sum_m
c_{(s/2)}(m)q^m\ ,\ s\neq 0$,\ $\Phi_{-n}(q,1):=\sum_m c(m) q^m$ 
and $\cK={8\pi\o 3\sqrt
3}e^{1-\gamma_E}$.
This gives precisely the integrals $\tilde\cI, \cI$ given in eq. (A.31) 
and (A.47) of \HM.\br

$(ii)$ $F^4$ in $d=8$, with
$\Phi_{-4}(q,y)={E_4J_{E_8}(q,y)\o\eta^{24}}$:
$$
\eqalign{
\Delta_{(\tr F_{E_8}^2)^2}
\!\!\!\!\!\!\!\!\!\!\!\!\!\!\!\!\!\!
&\qquad\ \ \ =
-c_{(2)}(0)\ln({\cal K}T_2U_2)\cr
&+4\re\lf\{\!\!\sum_{(k,l)>0}c_{(2)}(kl)\Li_1(x)-{6\o \pi
T_2U_2}c_{(1)}(kl)
\lf[(kT_2+lU_2)\Li_2(x)+{1\o 2\pi}\Li_3(x)\ri]\ri.\cr
&+\lf.{9\o \pi^2T_2^2U_2^2}c(kl)\lf[(kT_2+lU_2)^2\Li_3(x)+
{3\o 2\pi}(kT_2+lU_2)\Li_4(x)+{3\o 4\pi^2}\Li_5(x)\ri]\ri\}\cr
&-{2\pi c_{(1)}(0)\o 15}{U_2^2\o T_2}-{6c_{(1)}(0)\zeta(3)\o
\pi^2T_2U_2}+
{4\pi c(0)\o 105}{U_2^3\o T_2^2}+{27c(0)\zeta(5)\o 2\pi^4T_2^2U_2^2}+
{\pi\o 3}c_{(2)}(0)U_2+384\pi T_2\ .\cr}
\eqn\Eight
$$
This gives precisely the integrals given in eq.\ (E.27) of \KO, which
we need
in section 2.1.

The correction $\Delta_{\tr F_{E_8}^2\tr F_{E'_8}^2}$, which we also
need in section 2.1, is easily obtained from \Eight\ by replacing the
coefficients $c_{(s/2)}(n)$ 
with\foot{The last term becomes $-192\pi T_2$.}:
$$
\eqalign{
\sum c(m)q^m&={J_{E_8}(q,1)^2\o\eta^{24}}\cr
\sum c_{(1)}(m)q^m&=
{3\o 4}(y_1{\p\o\p y_1})^2{J_{E_8}(q,y_1)J_{E_8}(q,y_2)
\o\eta^{24}}+
{3\o 4}(y_2{\p\o\p y_2})^2{J_{E_8}(q,y_1)J_{E_8}(q,y_2)
\o\eta^{24}}\lf.\ri|_{z_i=0}\cr
\sum c_{(2)}(m)q^m&={9\o 4}(y_1{\p\o\p y_1})^2(y_2{\p\o\p y_2})^2
{J_{E_8}(q,y_1)J_{E_8}(q,y_2)\o\eta^{24}}\lf.\ri|_{z_i=0}\ .\cr}
\eqn\GGSS
$$
We see that the (harmonic) $\Li_1$--term arises from maximally
differentiating the Jacobi function $\Phi_{-n}(q,y)$, i.e., its
coefficients $c_{(n/2)}(kl)$ involve powers of $E_2^{n/2}$.  On the
other hand, for the maximally non--harmonic terms (proportional to
${1\o (T_2U_2)^{n/2}}$) the coefficients $c(kl)$ of $\Phi_{-n}(q,0)$
appear. In fact, the expressions in the brackets [\ ] are precisely the
Bloch--Ramakrishnan--Wigner polylogarithms \doubref\gm\kawainew.

\section{Gauge groups $G\times G'\subset E_8\times E_8'$}

\def\t{{1\o 3}}

Threshold corrections for gauge groups $G\times G' \subset E_8 \times
E_8'$ are obtained by introducing Wilson lines. We consider
two cases: $(I)$ $[SU(2)\times E_7]^2$ and $(II)$ $[SU(3)\times
E_6]^2$,
for which appropriate (discrete) Wilson lines are:
$$
\eqalign{
(I) & \ \  a_1^I=\h(1,-1,0,0,0,0,0,0), \ \h(1,-1,0,0,0,0,0,0)\ ,\cr
(II)& \ \ a_1^I=\t(1,1,-2,0,0,0,0,0), \ \t(1,1,-2,0,0,0,0,0)\ .\cr}
\eqn\wlEs
$$
The internal part of the partition function $Z_{(18,2)}(q,\ov q)$ \gins\
becomes  a $\ZZ_M$ orbifold with K\"ahler modulus $\tilde T=MT$
and complex structure modulus $\tilde U=U/M$  (where $M=2$ and $M=3$,
respectively):
$$
Z_{(18,2)}^{G\times G'}(q,\ov q)=
\sum_{(h,g)}\sum_{{m_1,m_2 \atop n_1+{h\o M},n_2}}e^{{2\pi i\o M}gm_1}
q^{\h \tilde P_L^2}\ov q^{\h \tilde P_R^2}\ \cC_{(h,g)}(q)\ ,
\eqn\partE
$$
It is shifted by
$\th={1\o M}(0,0,1,0)$ in $(P_L,P_R)\in  {\cal N}_{2,2}$ and
$\Theta=a_1$ in $E_8\times E_8$ with:
$$
\eqalign{
\cC_0:=\cC_{(0,0)}(q)&={E_4^2\o\eta^{24}}=\eta^{-24}(
Z_{E_7^0}Z_{A_1^0}+Z_{E_7^1}Z_{A_1^1})^2\ ,\cr
\cC_1:=\cC_{(0,1)}(q)&={1\o
4}\eta^{-24}(\theta_4^2\theta_3^6+\theta_3^2\theta_4^6)^2
=\eta^{-24}(Z_{E_7^0}Z_{A_1^0}-Z_{E_7^1}Z_{A_1^1})^2\ .\cr}
\eqn\EF
$$
We have introduced here
the lattice partition functions for $E_7$ \except\
and $A_1$:
$$
\eqalign{
Z_{E_7^0}&=
\theta_3^7(2\tau)+7\theta_3^3(2\tau)\theta_2^4(2\tau)\cr
Z_{E_7^1}&=
\theta_2^7(2\tau)+7\theta_2^3(2\tau)\theta_3^4(2\tau)\cr
Z_{A_1^0}&=\theta_3(2\tau)\cr
Z_{A_1^1}&=\theta_2(2\tau)\ .\cr}
\eqn\partdec
$$
The twisted sector functions follow from modular invariance.
Similarly, for the $E_6$ model we get:
$$
\eqalign{
\cC_0:=\cC_{(1,1)}(q)&={E_4^2\o\eta^{24}}=\eta^{-24}(Z_{E_6^0}
Z_{A_2^0}+2
Z_{E_6^1} Z_{A_2^1})^2 \cr
\cC_1:=\cC_{(1,\theta)}(q)&=\cC_{(1,\theta^2)}(q)=\eta^{-24}(Z_{E_6^0}
Z_{A_2^0}-Z_{E_6^1} Z_{A_2^1})^2\ ,\cr}
\eqn\charesix
$$
where we have introduced the following $E_6$
\except\ and $A_2$--characters:
$$
\eqalign{
Z_{E_6^0}&=\h\lf\{\theta_3(3\tau)\theta_3(\tau)^5+\theta_4(3\tau)
\theta_4(\tau)^5+\theta_2(3\tau)\theta_2(\tau)^5\ri\}\cr
Z_{E_6^1}&=\h\lf\{\theta\lf[{1/3 \atop 0}\ri](3\tau)\theta_2(\tau)^5+
\theta\lf[{4/3 \atop 0}\ri](3\tau)\theta_3(\tau)^5-\rho^\h
\theta\lf[{4/3 \atop 1}\ri](3\tau)\theta_4(\tau)^5\ri\}\cr
Z_{E_6^{\ov 1}}&=\h\lf\{\theta\lf[{5/3 \atop
0}\ri](3\tau)\theta_2(\tau)^5+
\theta\lf[{2/3 \atop 0}\ri](3\tau)\theta_3(\tau)^5-\rho
\theta\lf[{2/3 \atop 1}\ri](3\tau)\theta_4(\tau)^5\ri\}\cr
Z_{A_2^0}&=\theta_3(2\tau)\theta_3(6\tau)+
\theta_2(2\tau)\theta_2(6\tau)\cr
Z_{A_2^1}&=\theta_3(2\tau)\theta\lf[{4/3\atop
0}\ri](6\tau)+\theta_2(2\tau)
\theta\lf[{1/3\atop 0}\ri](6\tau).\cr }
\eqn\partdecesix
$$
Again, the twisted sector functions follow from modular invariance.
The dependence on the skew eigenvalues of $F$ may be easily introduced
for each sector by replacing the $\theta$--functions with Jacobi
functions (B.29):
$$
\eqalign{
(I)\ \ \ &J_{E_7,i}(q,y_1,y_2)=
Z_{E_7^0}(q,y_1)Z_{A_1^0}(q,y_2)-Z_{E_7^1}(q,y_1)Z_{A_1^1}(q,y_2)\cr
(II)\ \ \ &J_{E_6,i}(q,y_1,y_2)=Z_{E_6^0}(q,y_1)Z_{A_2^0}(q,y_2)-
Z_{E_6^1}(q,y_1)Z_{A_2^1}(q,y_2)\ ,\cr}
\eqn\jacs
$$
for the coset $i=1$. For the subsequent
world--sheet $\tau$--integration, it is convenient to express
the orbifold sector sum in \partE\ as sum over the cosets \msi
$$
Z(q,\ov q,\tilde T,\tilde U)_i=\nu_i\sum_{A_i}\ q^{\h |\tilde P_L|^2}\
\ov q^{\h |\tilde P_L|^2}\ ,\ i=1,\ldots,M+1\ ,
\eqn\cosets
$$
with the $A_1=\{m_1\in M\ZZ; m_2,n_1,n_2 \in \ZZ\}$,
$A_2=\{n_1\in {\ZZ\o M}; m_1,m_2,n_2 \in \ZZ\}$ etc. and
$\nu_i=vol({\cN}_{2,2_i})
=\{1,{1\o M},\ldots,{1\o M}\}$.
The function $\tau_2
Z(q,\ov q,\tilde T,\tilde U)_1$ is invariant under $\Ga_0(M)_\tau
\times\Ga^0(M)_{\tilde T}\times\Ga_0(M)_{\tilde U}$.

After expressing the $G\times G'$ currents as $E_8\times E_8$
currents, we follow \ellis\ to extract the relevant gauge
contractions:
$$
\eqalign{
\Delta_{\tr F^2_\alpha\tr F^2_\beta}&=
\int {d^2\tau \o \tau_2}\lf\{a [Z(q, \ov q,\tilde T,\tilde
U)_0-1]+{1\o (2\pi i)^4}{\p^4\o\p \ov z_\alpha^2 \p \ov z_\beta^2}\cr
&\times \int {d^2\tau \o \tau_2}\sum_{i=1}^{M+1}
b[Z(q, \ov q,\tilde T,\tilde U)_i{\tilde J_{G,i}(\ov q,\ov y_1,\ov
y_2)\tilde J_{G',i}(\ov q,\ov y_3,\ov
y_4)\o\ov\eta^{24}}-\nu_ib_i(0)]\lf.\ri|_{z_i=0}\ ,\cr}
\eqn\semifinal
$$
with $b_i={\p^4\o\p y_\alpha^2 \p
y_\beta^2}{J_{G,i}J_{G',i}\o\eta^{24}}\lf.\ri|_{z_i=0}=\sum_k
b(k)_iq^k$. This expression is the generalization of \todo\ to
subgroups $G\times G'\subset E_8\times E_8'$. We have displayed the
coefficients  $(\alpha,\beta,b_1,b_2)$ in the following tables, next to
two additional numbers $c,\tilde b$,  which will prove to be useful
later to write down the final result in a closed form:

\vskip 1cm
\goodbreak
{\vbox{\ninepoint{  $$
\vbox{\offinterlineskip\tabskip=0pt \halign{\strut\vrule# &~~$#$~~\hfil
&\vrule# &~~$#$~~\hfil &~~$#$~~\hfil &~~$#$~~\hfil &~~$#$~~\hfil
&~~$#$~~\hfil &~~$#$~~\hfil &\vrule# &~~$#$~~\hfil &\vrule# \cr
\noalign{\hrule} & \tr F^2_\alpha\tr F^2_\beta && a & b & b_1 & b_2 & c
& \tilde b  & \cr \noalign{\hrule} & \tr F^2_{E_7}\tr F^2_{E_7'} && 0 &
{2\o 49} & 0 & 0 & 0  & {2\o 49} & \cr \noalign{\hrule} & \tr
F^2_{A_1}\tr F^2_{A_1'} && 0 & 2 & 0 & 0 & 0  & 2 & \cr
\noalign{\hrule} & {\tr F^2_{A_1}\tr F^2_{E_7'}} && -6 & {2\o 7} & 0 &
21 & 0  & {2\o 7} &\cr \noalign{\hrule} & \tr F^2_{A_1}\tr F^2_{E_7} &&
6 & {2\o 7} & -21 & 0 & -6  & {2\o 7} & \cr \noalign{\hrule} & \tr
F^2_{E_7}\tr F^2_{E_7} && 0 & {2\o 147} & 63 & 63 & 2  & {2\o 49} & \cr
\noalign{\hrule} & \tr F^2_{A_1}\tr F^2_{A_1} && 0 & 2 & -5 & 7 & \  &
\  & \cr \noalign{\hrule}} \hrule}$$ \vskip-10pt \noindent{\bf Table
2:} \sl Coefficients $(a,b,b_i)$ in \semifinal\ for heterotic $F^4$
corrections with gauge group $[E_7\times SU(2)]^2$. In addition,
$\Delta_{\tr F_{A_1}^4}=-{1\o 3}\Delta_{(\tr F_{A_1}^2)^2}$ and
$\Delta_{\tr F_{E_7}^4}=0$.} \vskip10pt}}

\vskip 1cm
\goodbreak
{\vbox{\ninepoint{  $$
\vbox{\offinterlineskip\tabskip=0pt \halign{\strut\vrule# &~~$#$~~\hfil
&\vrule# &~~$#$~~\hfil &~~$#$~~\hfil &~~$#$~~\hfil &~~$#$~~\hfil
&~~$#$~~\hfil &~~$#$~~\hfil &\vrule# &~~$#$~~\hfil &\vrule# \cr
\noalign{\hrule} & \tr F^2_\alpha\tr F^2_\beta && a & b & b_1 & b_2 & c
 & \tilde b & \cr \noalign{\hrule} & \tr F^2_{E_6}\tr F^2_{E_6'} && 0 &
{1\o 8} & 0 & 0 & 0  & {1\o 18} & \cr \noalign{\hrule} & \tr
F^2_{A_2}\tr F^2_{A_2'} && 0 & \h & 0 & 0 & 0  & \h & \cr
\noalign{\hrule} & {\tr F^2_{A_2}\tr F^2_{E_6'}} && -3 & {1\o 6} & 0 &
18 & 0  & {1\o 6} & \cr \noalign{\hrule} & \tr F^2_{A_2}\tr F^2_{E_6}
&& 3 & {1\o 6} & -18 & 0 & -3  & {1\o 6} & \cr \noalign{\hrule} & \tr
F^2_{E_6}\tr F^2_{E_6} && 0 & {1\o 24} & 36 & 36 & {3\o 2}  & {1\o 18}
& \cr \noalign{\hrule} & \tr F^2_{A_2}\tr F^2_{A_2} && 0 & {3\o 4} & -3
& 9 & \   & \  & \cr \noalign{\hrule}} \hrule}$$ \vskip-10pt
\noindent{\bf Table 3:} \sl Coefficients $(a,b,b_i)$ in \semifinal\ 
for heterotic $F^4$ corrections with  gauge group $[E_6\times
SU(3)]^2$. In addition, $\Delta_{\tr F_{A_2}^4}=-{1\o 3}\Delta_{(\tr
F_{A_2}^2)^2}$ and $\Delta_{\tr F_{E_6}^4}=0$.} \vskip10pt}}

The techniques to perform world--sheet torus integrals over Narain
coset sums
$$
\Delta={1\o (2\pi i)^n}{\p^n\o \p\ov z^n}
\int{d^2\tau\o \tau_2}\sum_{i=1}^{M+1}\ [Z(q,\ov q,T,U)_i
\tilde \Phi_{-n,i}(\ov q,\ov y)-\nu_i c_{(n/2),i}(0)]\lf.\ri|_{z=0}
\eqn\basicset
$$
have been developed in \msi\ and extended\foot{Later,  in \mario\ also
integrals over coset sums have been calculated by an  independent
method. These results completely agree with our findings in \LS.} in
\LS. Essentially, $\Delta$ integrates to a sum over two sectors:
$$
\kern-1.5em
\eqalign{\Delta
&={1\o (2\pi i)^n}{\p^n \o \p z^n}\ \times\cr
&\lf\{\Big[\sum_b\!\sum_{k>0\atop l\in\ZZ}
\sum_{p>0}{2\o\sqrt{p^2-{mz^2\o MT_2U_2}}}\,
e^{-2\pi(kT_2+lMU_2)\sqrt{p^2-{mz^2\o MT_2U_2}}}e^{2\pi
ip(kT_1+lMU_1)}\
c_1(Mkl,b)y^b\ri.\cr
&+{2\o\sqrt{p^2-{mMz^2\o T_2U_2}}}\,
e^{-2\pi(k{T_2\o M}+lU_2)\sqrt{p^2-{mMz^2\o T_2U_2}}}
e^{2\pi ip(k{T_1\o M}+lU_1)}\ c_2({kl\o M},b)y^b\cr
&+\sum_{l>0}\sum_{p>0}{2\o\sqrt{p^2-{mz^2\o T_2U_2}}}\,
e^{-2\pi lMU_2\sqrt{p^2-{mz^2\o T_2U_2}}}e^{2\pi ip lMU_1}\ c_1(0,b)y^b\ri.\cr
&+\lf.{2\o\sqrt{p^2-{mz^2\o T_2U_2}}}e^{-2\pi lU_2\sqrt{p^2-{mz^2\o T_2U_2}}}
e^{2\pi iplU_1}\ c_2(0,b)y^b+hc.\Big]\cr
&\!+\!\sum_b
\lf[{MU_2\o \pi}\sum_{j>0}{2\o\sqrt{j^2-mz^2M^2{U_2\o T_2}}}+
\lf({2\o\sqrt{j^2-{mz^2\o T_2U_2}}}-{2\o\sqrt{j^2-{mz^2\o T_2U_2}+
{\epsilon\o \pi T_2U_2}}}\ri)\ri]c_1(0,b)y^b\cr
}
$$
\goodbreak\vskip -.5cm
$$
\eqalign{
&\lf.\lf.+\sum_b\lf[{U_2\o \pi}\sum_{j>0}{2\o\sqrt{j^2-mz^2{U_2\o
T_2}}}
+\lf({2\o\sqrt{j^2-{mz^2\o
T_2U_2}}}- {2\o\sqrt{j^2-{mz^2\o T_2U_2}+
{\epsilon\o \pi T_2U_2}}}\ri)\ri]c_2(0,b)y^b\lf.\ri\}\lf.\cr
&+\lf.{\pi\o 3}{T_2\o M}\sum_{i=1}^{M+1}\sum_{s=0}^{n/2}{1\o
s+1}E_2^{s+1}
F_{i,s}\ri|_{{\rm coeff}(q^0)}
\cr& -(c_{(n/2),1}(0)+c_{(n/2),2}(0))
[\ln\epsilon+\gamma_E+1+\ln({2\o 3\sqrt 3})]\,\ri|_{z=0}\ ,}
\eqn\RESULTtwo
$$
with ${3\o 2^{n/2}}\lf.(y {\p \o \p y})^n\tilde
\Phi_i(q,y)\ri|_{z=0}=:\sum\limits_{s=0}^{n/2} \hat{E_2}^s F_{i,s}$.
Similar as for the $E_8$ model \dif, we need for case $(i)$:
$$\eqalign{
(y{\p \o \p y_1})^2J_{E_7,1}(q,y_1,y_2)\attac{z_i=0}\!\!\!\!\!
&={7\o 48}\theta_3^2\theta_4^2
[E_2(\theta_3^4+\theta_4^4)+\theta_2^8-2\theta_3^4\theta_4^4]\ ,\cr
(y{\p \o \p y_2})^2J_{E_7,1}(q,y_1,y_2)\attac{z_i=0}\!\!\!\!\!
&={1 \o 48}\theta_3^2\theta_4^2[
E_2(\theta_3^4+\theta_4^4)-3\theta_2^8-2\theta_3^8-2\theta_4^8
+2\theta_3^4\theta_4^4]\ ,\cr}
\eqn\Stephan
$$
and for case $(ii)$:
$$\eqalign{
(y{\p \o \p y_1})^2J_{E_6,1}(q,y_1,y_2)\attac{z_i=0}\!\!\!\!\!
&={1 \o 4}f_0(f_0-3f_3-E_2f_1)\
,\cr
(y{\p \o \p y_2})^2J_{E_6,1}(q,y_1,y_2)\attac{z_i=0}\!\!\!\!\!
&={1 \o 12}f_0(f_0+9f_3-E_2f_1)\
,\cr}
\eqn\Nick
$$
with $f_1=Z_{A_2^0},\ f_0=\lf({\eta^3(\tau)\o \eta(3\tau)}\ri)^3$
and $f_3=\lf({3\eta^3(3\tau)\o \eta(\tau)}\ri)^3$.

While we need in the present paper only the harmonic pieces of the
threshold corrections, it may nevertheless be instructive to the reader
to note how easily also the non--harmonic terms derive from our
generating formulae \RESULT\ and \RESULTtwo. Evaluating the harmonic
part of \semifinal, we then arrive at our final result
(after dropping the pieces linear in $T_2$, which may be easily
derived from \Stephan\ and \Nick):
$$\eqalign{
\Delta^{\rm harmonic}_{\tr F_\alpha^2 \tr F_\beta^2}&=
4\re\lf\{\!-a\ln \eta(\tilde T)\eta(\tilde U)
-c\tilde b \ln\eta({\tilde T \o M})\eta(\tilde U)
-c\tilde b\ln \eta(\tilde T)\eta(M\tilde U)\cr
&\lf.+\tilde b\!\! \sum_{(k,l)>0}\! b^{\alpha\beta}_{1}(Mkl)
\Li_1(e^{2\pi i(k\tilde T +lM\tilde U)})
+\tilde b\!\! \sum_{(k,l)>0}\! b^{\alpha\beta}_{2}({kl \o M})
\Li_1(e^{2\pi i(k {\tilde T\o M} +l\tilde U)})\ri\}, \cr}
\eqn\last
$$
with the coefficients $b^{\alpha\beta}_{i}$ defined by:
$$
{1\o \eta(q)^{24}}
(y{\p \o \p y_\alpha})^2J_{G,i}(q,y_1,y_2)
 (y{\p \o \p y_\beta})^2J_{G',i}(q,y_1,y_2)\lf.\ri|_{z_i=0}=
\sum_m b^{\alpha\beta}_{i}(m)q^m\ .
\eqn\expan
$$
analogous to \GGSS.

In order to facilitate the comparison with the geometrical formulae of
section 2.3, we present here the first terms of the asymptotic
$q$-series  of the corrections \last\ for the $E_6$ model (which indeed
coincide with the $q$ expansions of \esix). In fact, the geometrical
couplings were defined at a fixed value of $U$, and it is not entirely
trivial to evaluate \last\ at this value.  Explicitly, for the $E_6$
model, where $U={\bf T}^{-1}\cdot\rho=\rho-1$ \LS, we find the
following expansions:
\def\frac{\Coeff}
\def\Mfunction{{\cal O}}
$$
\eqalign{
\Delta_{E_6{E_6}'}(T)\ &=\
- {\frac{1}{3}} \log (q) + 6\,q + 14\,{q^3} -
  {\frac{33\,{q^4}}{2}} + {{\Mfunction(q)}^{5}}
\cr
\Delta_{E_6{A_2}}(T)\ &=\
{\frac1{2}}{\log (q)} - 2\,q + 15\,{q^2} -
  {\frac{110\,{q^3}}{3}} + {\frac{263\,{q^4}}{2}} +
  {{\Mfunction(q)}^{5}}
\cr
\Delta_{E_6{A_2}'}(T)\ &=\
{\frac{2}{3}}\,\log (q) + 18\,{q^2} -
  36\,{q^3} + 135\,{q^4} + {{\Mfunction(q)}^{5}}
\cr
\Delta_{A_2{A_2}'}(T)\ &=\
 - {\frac{1}{3}}\log (q) + 24\,q - 81\,{q^2} +
  392\,{q^3} - 1848\,{q^4} + {{\Mfunction(q)}^{5}}
\ .\cr
}\eqn\qexpan
$$
Moreover, the solutions of the inhomogenous PF equation \sourceN\ for
$N=3$ look:
$$
\eqalign{
\mu_1(T)\ &=\
108\,q - 486\,{q^2} + 2268\,{q^3} - 10989\,{q^4} +
  {{\Mfunction(q)}^5}
\cr
\mu_2(T)\ &=\
108\,q - 810\,{q^2} + 4572\,{q^3} - 24597\,{q^4} +
  {{\Mfunction(q)}^5}
\ .}\eqn\muexpan
$$

\section{Jacobi functions}

A Jacobi form (for more details see \zagier) $f_{s,m}$ of weight
$s$ and index $m$ enjoys
$$\eqalign{
f_{s,m}\lf({a\tau +b \o c\tau+d},{z\o c\tau+d}\ri)&=(c\tau+d)^s
e^{2\pi i {m c z^2 \o c\tau+d}}f_{s,m}(\tau,z)\ ,\cr
f_{s,m}(\tau,z+\lambda\tau+\mu)&=e^{-2\pi im(\lambda^2\tau+2\lambda z)}
f_{s,m}(\tau,z)\ ,\cr}
\eqn\enjoy
$$
for $\lf(\matrix{a&b\cr c& d\cr}\ri)\in SL(2,\ZZ)$ and
$\lambda,\mu\in\ZZ$.
With
$$
\theta\lf[{\alpha\atop\beta}\ri](q,y)=\sum_{n\in \ZZ}
q^{\h(n+\h \alpha)^2}e^{\pi i(n+\h\alpha)\beta}\ y^{n+\h \alpha}
\eqn\jac
$$
the function
$$
J_{E_8}(q,y):=\h\sum_{(\alpha,\beta)}
\theta\lf[{\alpha\atop\beta}\ri](q,y)^8=
1+q(126 +56 y^{-2}+56 y^2+y^{-4}+y^{4})+\ldots
\eqn\jac
$$
is a Jacobi function of weight $4$ and index $m=4$, whereas
$$
E_{4,1}(q,y)=\h[\theta_2(q,y)^2\theta_2^6+
\theta_3(q,y)^2\theta_3^6+\theta_4(q,y)^2\theta_4^6]\ ,
\eqn\eisen
$$
has index $m=1$
($J_{E_8}(q,y)=E_{4,1}(q,y^2)$).
We use the notation $\theta_1=\theta\lf[{1\atop 1}\ri],\
\theta_2=\theta\lf[{1\atop 0}\ri],\
\theta_3=\theta\lf[{0\atop 0}\ri]$ and $\theta_4=\theta\lf[{0\atop
1}\ri]$.


\listrefs
\end